\documentclass[letter,12pt]{article}
\usepackage[dvips]{graphicx}
\usepackage{graphics}
\usepackage{psfig}
\usepackage{textcomp}
\usepackage{url}
\usepackage{amsfonts}
\usepackage{mathenv}
\usepackage{tipa}

\begin{document}


\title
{\bf {How do clarinet players adjust the
resonances of their vocal tracts for different playing effects? 
}}

\author{\Large Claudia Fritz \footnote{Current address: LIMSI-CNRS, BP 133, 91403 Orsay, France; Electronic mail: claudia.fritz@ens-lyon.org}\ \ \  and Joe Wolfe\\
\date{\rmfamily\itshape UNSW, School of Physics, NSW 2052 Sydney, Australia}}

\maketitle

\abstract 

 In a simple model, the reed of the clarinet is mechanically loaded by
 the series combination of the acoustical impedances of the instrument
 itself and of the player's airway. Here we measure the complex
 impedance spectrum of players' airways using an impedance head
 adapted to fit inside a clarinet mouthpiece. A direct current shunt
 with high acoustical resistance allows players to blow normally, so
 the players can simulate the tract condition under playing
 conditions. The reproducibility of the results suggest that the
 players' ``muscle memory" is reliable for this task. Most players use
 a single, highly stable vocal tract configuration over most of the
 playing range, except for the altissimo register. However, this
 ``normal'' configuration varies substantially among musicians. All
 musicians change the configuration, often drastically for ``special
 effects'' such as glissandi and slurs: the tongue is lowered and the
 impedance magnitude reduced when the player intends to lower the
 pitch or to slur downwards, and vice versa.

   PACS:43.75.Pq, 43.75.Yy, 43.75.St, 43.58.Bh

\section{INTRODUCTION} 
\label{sec:intro} 
Acousticians (Backus~\cite{Bac85}, Benade~\cite{Ben83},
Hoekje~\cite{Hoe86}, Johnson et
al.\cite{Joh86}, Wilson~\cite{Wil96}) are divided over the extent of
the influence of the 
respiratory tract in playing reed instruments, of which the clarinet
is the most studied example. The reed and the airflow past it interact
with acoustical waves 
in the bore of the instrument and with waves in the player's tract. A
simple argument shows that the acoustical impedances of these are
approximately in series\cite{Ben83}. The cross section of the clarinet
bore is rather smaller than that of the tract, so its characteristic
impedance is higher. Further, the resonances in the instrument have a
high value of quality factor, so the peaks in impedance have high
value and, to first order, usually determine the playing regime of the
bore-reed-tract system~\cite{Bac61}. Nevertheless, the effects of the
impedance spectrum of the vocal tract, even if smaller than those of
the clarinet, may be important, because musicians are often interested
in subtle effects. For instance, a 1$\%$ change in frequency could be
a large mistuning for a musician, and subtle changes in the spectral
envelope may be important in controlling timbre and musical
expression. Most researchers agree that the effect is small but
important - even if they do 
not necessarily agree about how the vocal tract affects the sound
production - except for Backus who considers that the player's tract has a
negligible influence on the instrument tone.

Strictly speaking, it is the impedance of the entire
airway of the player, from mouth to lungs, that loads the reed or lips
and that drives the air flow past them. However, Mukai~\cite{Muk92}
reported that 
experienced players of wind instruments keep the glottis (the aperture
between the vocal folds) almost closed when playing. This is very
important to the possible influence of the tract: with an open
glottis, the airway has relatively weak resonances, because it is
terminated with the high losses in the lungs and lower airways. In
contrast, an almost closed glottis provides a high coefficient of
reflection for acoustic waves at all but the lowest frequencies, and
so would be expected to give strong resonances, similar to those that
give rise to formants in speech. For that reason, we shall refer
hereafter to the player's vocal tract as the resonator that is
controlled by the player.  

\paragraph{} 
Impedance measurements have
been made previously (Benade~\cite{Ben83}, Backus~\cite{Bac85},
Hoekje~\cite{Hoe86}\ and Wilson~\cite{Wil96}) but are not fully
exploitable or applicable due to the fact that they either were
performed under conditions that do not closely resemble those used to
play an instrument, or that they lacked phase information, or
contained high levels of background noise. Moreover, they were made in
most cases on only one subject. The measurement conditions
should reproduce, as much as possible, the playing condition, so that
the player can automatically adopt the tract configurations used in
playing under particular conditions. For example, Benade~\cite{Ben83}
measured the impedance of a clarinettist's tract by inserting into the
player's mouth a pipe containing the acoustical source and the
microphone. The pipe inner diameter was 20 mm, which
forces the player to open the mouth considerably more than he would
when playing a clarinet. This problem was solved by
Hoekje~\cite{Hoe86}, who used a similar arrangement, with the exception
that he reduced the size of the part which goes into the musician's
mouth. Overall, Hoekje measured somewhat low values of the impedance (maxima
about 8 MPa.s.m$^{-3}$), therefore much smaller
than the maxima measured for the clarinet using most fingerings (Wolfe
\emph{et al.}~\cite{webUNSW}, Backus~\cite{Bac74}). This may be explained by
the fact the player could not 
breathe into the apparatus, nor was the glottis aperture monitored. It
is likely, therefore, that the subject may
have relaxed the glottis, and thus reduced the magnitude of the airway
resonances, as discussed above. This is the case too with
Backus~\cite{Bac85}'s measurements which only give a maximum magnitude
of 5 MPa.s.m$^{-3}$. That would explain as well why he reported
that the values he obtained could not be consistently reproduced, as a
musician cannot be consistent with his glottis if he cannot
blow. Moreover, all these studies only give the amplitude of the
impedance but not the phase.  Wilson~\cite{Wil96} measured the complex
impedance in a situation in which a clarinettist 
could mime playing while exhaling, in order, to her opinion, to keep
the glottis open (so in contradiction with Mukai's cliches). These
measurements were made with a chirp of one third of a second 
duration, and so have a relatively high noise component. The
performers were three  
professional clarinettists and two advanced amateurs. She was also
able to obtain values for the impedance during playing at 
the frequencies of the harmonics of the note played. These are
interesting values but, because they are widely spaced in frequency, the
give little information about the tract configuration and few data for
numerical simulations. There may have been problems in consistency,
because the two methods did not always give similar results.

\paragraph{}
It has not yet proved possible to make accurate impedance measurements in the
vocal tract during playing because of the very high sound levels
produced by the reed. (The acoustic current produced in the tract by
the reed is comparable with that produced in the clarinet, so peak
pressure levels are high.) Consequently, it is still necessary to
measure clarinettists miming playing. In our measurements reported
here, a direct current shunt was placed in parallel with the impedance
head, to allow the players to blow normally, and so to adopt a tract
configuration approaching that used for playing.  

Our measurements were
done on professional clarinettists and advanced students. They were
asked to play notes on their own clarinet, set up for normal playing,
and then to mime playing on the instrument containing the impedance
head.  Notes over the range of the instrument were chosen, and players
were asked to play and to mime a range of conditions requiring
different embouchures to adjust the intonation or register, or to
produce other effects.

    \section{MATERIALS AND METHODS}  

\subsection{The impedance spectrometer} 

\begin{figure}[h!]   
\centering   
\includegraphics[width=12cm]{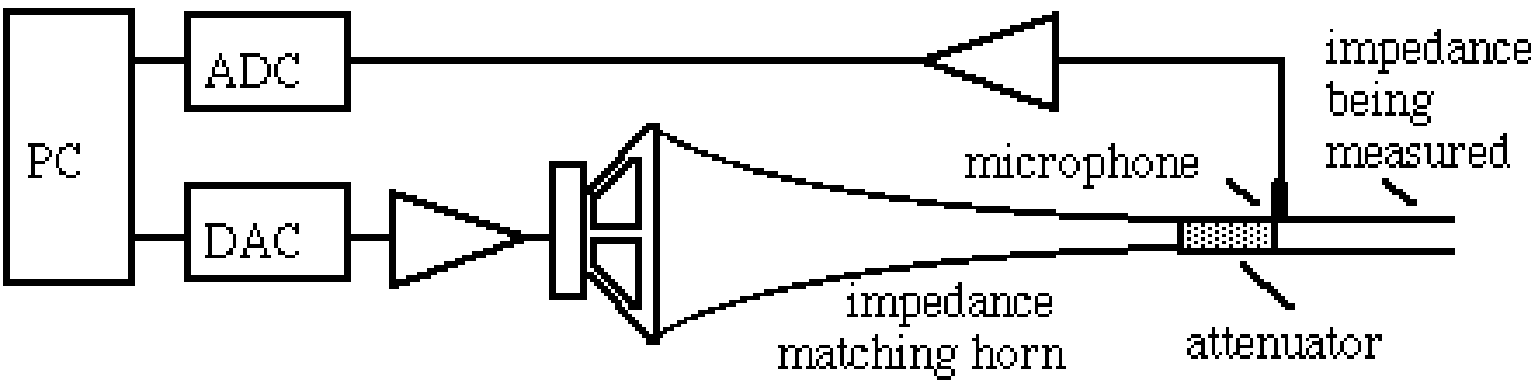}   
\caption{A schematic of the impedance spectrometer using the capillary method.}   
\label{fig:acuz} 
\end{figure}

 The setup is based on the impedance spectrometer developed previously
 \cite{Smi00}, which uses a source of acoustic current produced from
 an output with high acoustic impedance (see FIG.~\ref{fig:acuz})
 and which is calibrated using an acoustically infinite waveguide as
 the reference impedance. This reference is a straight, cylindrical
 stainless steel pipe, 7.8 mm in diameter and 42 m long, so that
 echos, in the frequency range of interest, return attenuated by about
 80 dB or more. Several compromises were made to incorporate an
 impedance head of this type into the clarinet mouthpiece so that it
 can measure the impedance that loads the clarinet reed without
 disturbing the player.

\begin{figure}[h!]   
\centering   
\includegraphics[height=3.5cm, width=8.7cm]{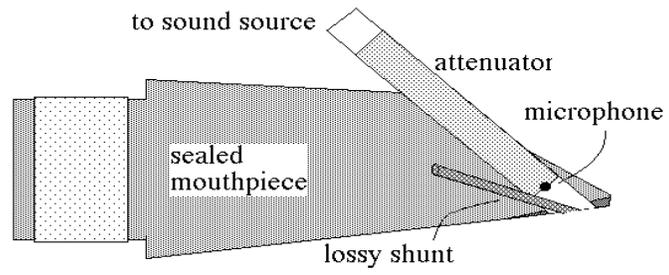}   
\caption{Cross section of the clarinet mouthpiece containing the
impedance head and a lossy shunt.}   
\label{fig:angle} 
\end{figure}

A range of impedance heads and (cylindrical) reference waveguides are
available. For this experiment, we chose to use one with diameter 7.8
mm, because it yields a cross sectional area comparable with that of
the effective surface area of the reed protruding past the lower lip
inside the mouth. Such an impedance head was mounted inside a modified
clarinet mouthpiece as is shown in FIG~\ref{fig:angle}. The angle is
chosen so that the head passes through the upper surface of the
mouthpiece just beyond the point where the player's teeth rest and
meets the lower surface at the position of the reed tip. The end of
the attenuator (the current source) and a small microphone (Countryman
CAI-B6 miniature B6, diameter 2 mm) positioned $l$ = 9 mm from the end
of the head, and the impedance at the end is calculated using the
transfer matrix for a cylindrical waveguide.

This angle produces an elliptical area at the end of the measurement
head. For calibration, this was simply sealed on the circular area of
the reference waveguide, with the centers aligned. Several other
geometries were also tried: one used a bent waveguide between the
measurement plane and the reference plane. Another used straight tubes
as here, but the extra volume at the ends of the ellipse were filled
with modelling compound. To estimate the effect of the discontinuities
thus produced, the impedance was measured for a range of waveguides
with simple, known geometries (cylindrical pipes of different
diameters and lengths), for which the impedance is known from other
measurements to agree well with theory.  The most successful fits were
obtained from the geometry shown: the simple straight impedance head
with the open elliptical end. For pipes of same diameter as the head,
the comparison between the measurement gives an error of 1\% in
frequency and up to 20\% in amplitude at high frequency.

The mouthpiece was sealed with epoxy so that the measurement head is
connected only to the player's tract and not to the clarinet. In any
case, the position of the head, which should measure the impedance in
the plane of the reed near its tip, prohibits the installation of a
reed. Preliminary experiments showed however that musicians could
reproduce embouchures that had very similar acoustic impedance
spectra. This suggested that they have a high developed sensory or
muscle memory and can mime easily a configuration that they use
regularly. This is not surprising: it is presumably what they must do
normally before playing in order to have the desired pitch and timbre
from the beginning of their first note. However, players are not
usually aware of the position of the vocal folds and the glottis and
so, if they are not blowing air, they may close them or relax them.
For that reason, a shunt with a DC impedance, judged by a clarinettist
to be comparable with that of a clarinet under normal playing
conditions, was introduced to allow subjects to blow normally.  A
small pipe (40 mm long and 3 mm diameter) was positioned to provide a
shunt or leak from the mouth to the outside air. Its short length
ensured that resonances and antiresonances fell beyond the frequency
range of interest and measurement, its diameter ensures that its
characteristic impedance is between 10 and 100 times larger than the
maxima in the vocal tract impedance with which it is in parallel, and
it was filled with acoustic wool which makes the impedance largely
resistive, reduces the turbulent noise due to flow and provides a DC
resistance comparable to that of a real clarinet.  

To prevent water
condensation in the measurement apparatus, a low voltage electrical
circuit was used to raise the temperature of the impedance head to
40\textdegree C.

\subsection{Procedure}  

Seventeen players took part in the experiment and their musical level
varied between advanced student and professional. They first answered
a survey about their musical backgrounds and their opinions about the
influence of the vocal tract when playing. Throughout all measurement
sessions, a digital audio tape recorder was used to record players
comments and played sounds. The microphone was positioned 10 cm from
the bell.  

For measurements, each player was asked first to play a note
\emph{mezzo forte} on his/her own clarinet, and then to mime playing
the same note on the modified clarinet. The notes, selected after
discussion with clarinettists, were (written) G3, G4, G5 and G6. G3 is
close to the lower end of the instrument range and uses almost the
full length of the nearly cylindrical part of the bore. It is a good
example of a note in the chalumeau register. G4 use the fundamental
mode of a relatively short section of the bore: it is an example of a
note in the throat register. G5 uses the speaker or primary register
key and the second resonance of a medium length tube: it is an example
of the clarion register. G6 uses two open register holes and is an
example of a note in the altissimo register.  

The subjects then played and mimed some unusual embouchures: some
peculiar configurations such as pitch bending (lowering the pitch
without changing the fingering), slurring a register change and
embouchures of their own suggestion used for different playing
conditions. They were also asked to mime embouchures described in
terms of vowels (in particular ``ee'' and ``aw''), a description used
by some clarinettists.  For the slurred register change, the musicians
were asked to mime over 5 seconds what they usually do less than a
second, during the transient between two notes.  

The measurements were made over the range 0.1-3 kHz, which includes
the playing range of the instrument.  In this range, there are usually
three vocal tract resonances, at typically 0.3, 1.3 and 2.3 kHz,
although the frequency varies among different players and playing
conditions. The sampling in the frequency domain was chosen as a
compromise between a high signal to noise ratio and precision in
frequency. The frequency resolution was set at 5.4 Hz. The measurement
time was set at 10 seconds (except for some unusual embouchures) as it
is tiring and hard for a musician to hold a constant embouchure
longer. 


\section{RESULTS}  

\subsection{The survey}  

Except from one amateur player, all the participating musicians
consider that their vocal tract has a very important influence on the
timbre. Regarding the pitch, four of them think that the vocal tract
is important whereas the thirteen others regard it as very important.  

For more specific details, we shall only quote here the musicians who
were the most able to describe their own utilisation of the vocal
tract. We shall retain their own vocabulary, which often corresponds
to mental and musical images. Some of the subjects, with busy
schedules as performing musicians, had done no teaching for many years
and were therefore not in the habit of describing what they do with
the mouth.   

Player B, a very experienced music teacher, reported having
reflected at depth on what she does in order to explain it to her
pupils. She changes the vocal tract shape for: 
\begin{itemize} 
\item note bending (i.e. adjusting the pitch using the mouth, rather
  than keys on the instrument); 
\item changing tonal colours to give character to interpretations. For
  that effect, she especially uses two configurations. In one, which
  she names for the vowel in ``hee'', she reports that she has the
  back and middle tongue in a high position, increased lip tension,
  the soft palate is lowered and the throat somewhat closed. This
  embouchure she uses and recommends for for brightening the sound. In
  another named for the vowel in ``haw'', she reports a high soft
  palate, the back of the tongue lowered and the throat more open.
  This she recommends and uses for darkening the timbre;
\item for changing articulation : the tongue has to be as close as
  possible to the tip of the reed to have a light articulation. So
  the ``hee'' configuration is usually more appropriate than the
  ``haw'' one. 
\end{itemize} 
Her tongue touches the lower lip but not usually the lower teeth. The
tongue can actually touch the lip/teeth in low or clarion register but
not in altissimo register. It is in general between 1 and 2 mm away
from the teeth.  

Player D, another experienced player and teacher reported lifting the soft
palate in order to obtain more resonance and projection which, she
said, induces a richer sound. She reports that her tongue  touches
neither the lower teeth nor the lower lip, and is in different
positions according to the register: 
\begin{itemize} 
\item for the low register, the tongue is low and arched, 1 cm away
  from the lip 
\item for the high register: the tongue is higher in the mouth, moves
  a little forward (about 8 mm away from the lip), becomes wider and
  flattens. 
\end{itemize}  

One advanced student, player H, prefers having the tongue high in the mouth so
the sound is more ``focused''. He uses changes in the vocal tract for
register change, large intervals, pitch bend and multiphonics. 

 Player C, a very experienced professional player, reported that he enriches
 the sound in high harmonics by opening the oral cavity. Further, he
 ``opens the throat'', but not necessarily the glottis, when he descends a
 register. Above all, however, he 
 reports using his facial muscles in order to modify the embouchure.  

Another very experienced professional player, player A, imagines, when playing,
``focussing the sound through the nose''. She has the impression that
the more her soft palate is arched the more the sound is
``focussed''. (It should be remarked that the velum must be closed or
very nearly closed during clarinet playing, to avoid a DC shunt
through the nose that would prohibit playing. However, the muscular
tension in the velum could in principle affect the impedance
spectrum.)    

In at least one case, disagreements among the opinions of the
musicians were reported. Player D reported that large mouth cavity was
useful for a ``rich, focussed'' sound, while others reported that they
achieved such a sound by lifting the tongue close to the soft
palate. One possible explanation is that the musicians in question have
different meanings for ``rich'' and especially for ``focussed'' in this
circumstance.

\subsection{Reproducibility of the impedance measurements}  

\begin{figure}[h!] 
\centering 
\includegraphics[width=7.5cm, height=5.5cm]{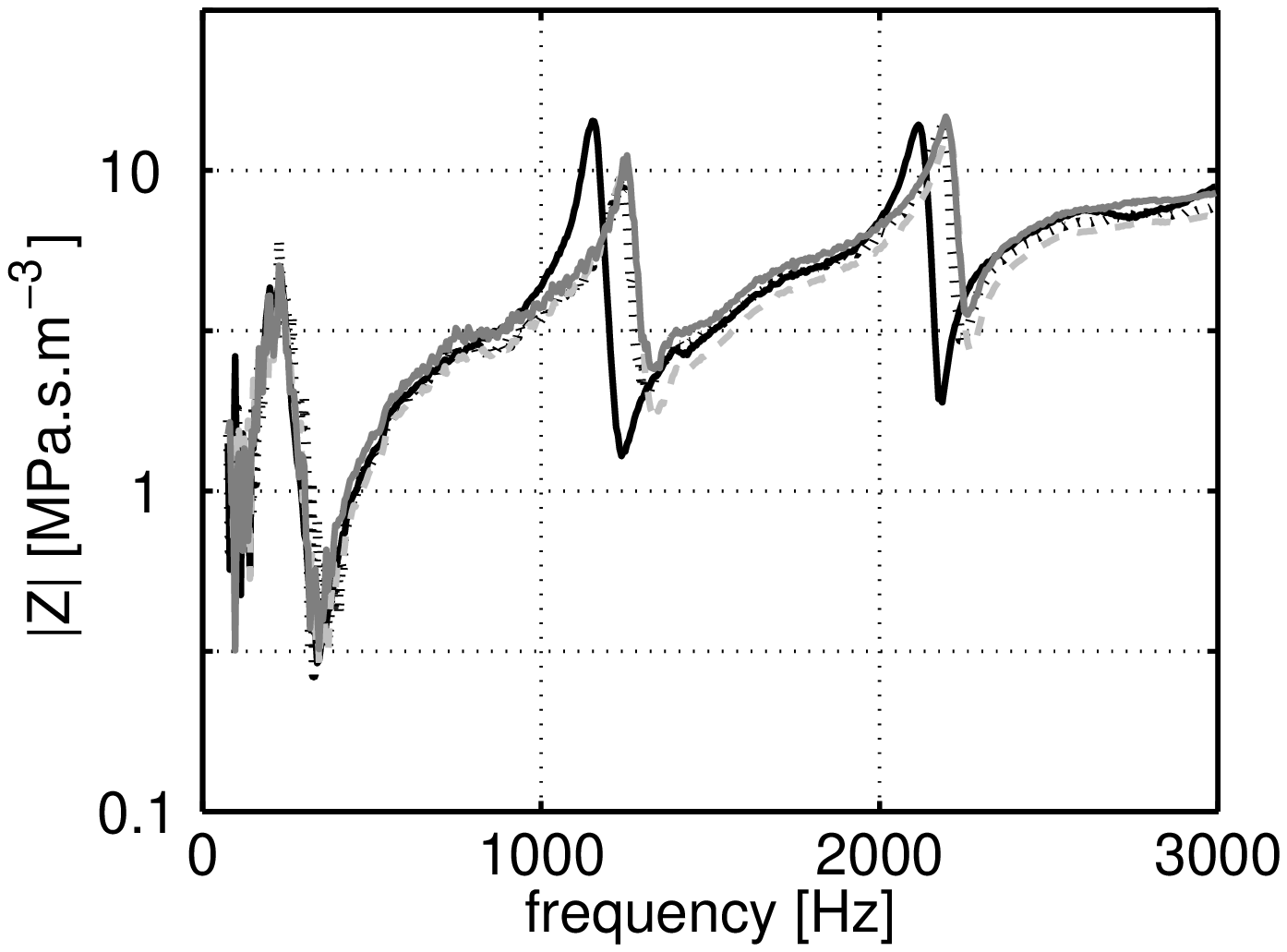} 
\includegraphics[width=7.5cm, height=5.5cm]{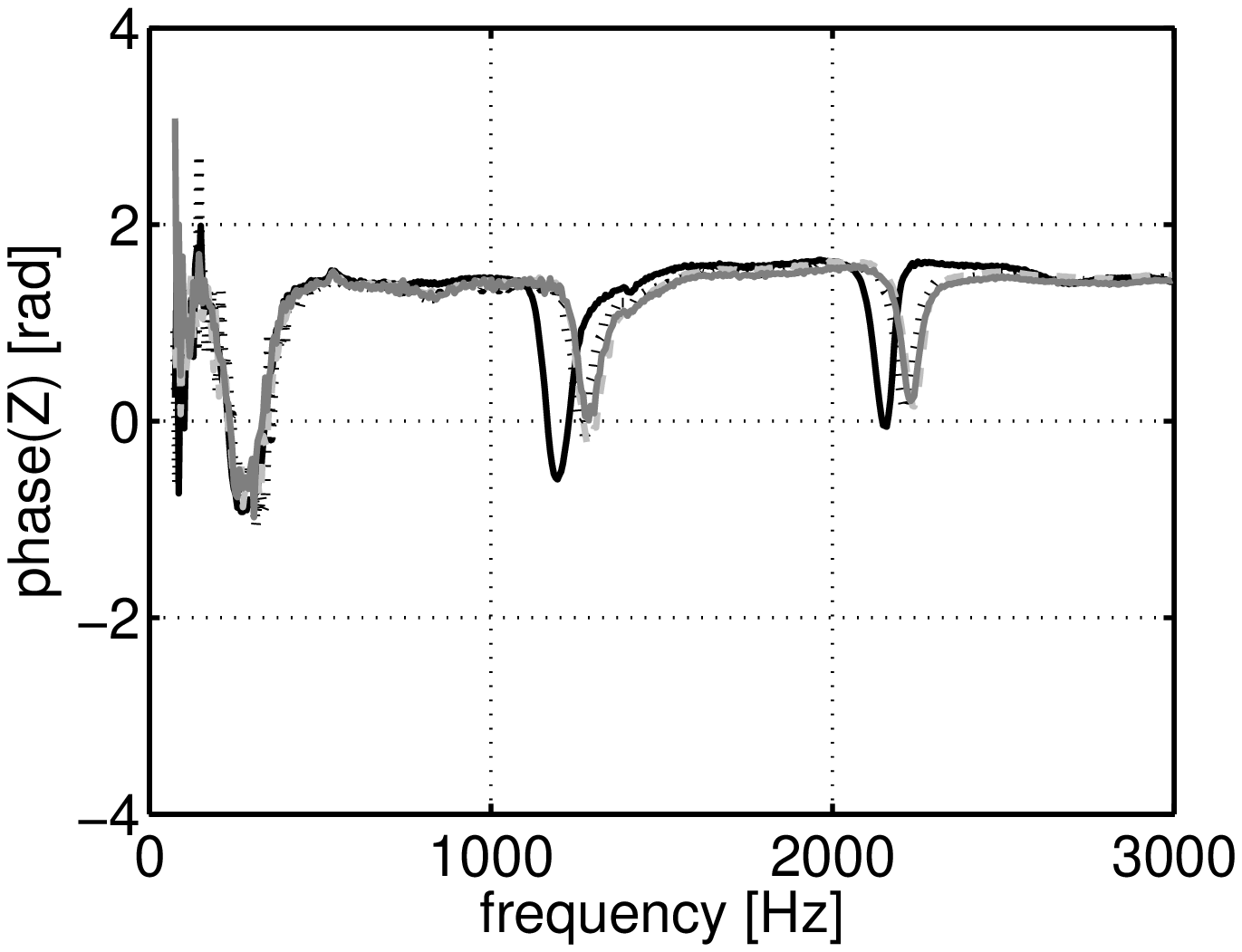}     
\caption{Typical results of measures testing the reproducibility of the
player's tract configuration: the impedance spectra of the vocal tract
  of the same player (player G) miming playing the note G3 on three
occasions over an interval of 40 minutes.}   \label{fig:reprod} 
\end{figure}

Reproducibility was tested on each musician by making about five
measurements of the embouchure for the same note (written G3) over the
course of a session (typically 40 minutes). Players were able to
repeat their embouchures rather reproducibly: in the typical result
shown in FIG.~\ref{fig:reprod}, the second resonance is obtained at
1250 Hz with a standard deviation of 3 \% in frequency and 15 \% in
amplitude.

\subsection{General comments}   

Most of the subjects in our study reported that, for normal playing,
they use an embouchure that varies little over most of the range,
except for the highest register. This was confirmed by the
measurements: for all players, the form of the impedance spectra is
quite stable over the whole register, except sometimes from the
altissimo register.  

The geometric average amplitude of the impedance is similar for all
musicians. The first peak, whose frequency is between 200 and 300 Hz,
has an amplitude between 1.8 and 5.6 MPa.s.m$^{-3}$ . The next
resonances are on the other hand different for both amplitude and
frequency. For some player embouchure combinations, the amplitudes are
in the range 30 to 100 MPa.s.m$^{-3}$  which is of the same order as that of the impedance of the clarinet at its resonances~\cite{webUNSW}.   

The difference between the impedance spectra recorded for the
``normal'' playing configuration and that measured for the tract
configuration used for ``special effects'' is not very large for any
of the student players measured. For some of the professional players,
however, the effect was very large. However, the spectra measured for
the different special effects also varied substantially among these
players, just as it did for normal playing.   

\begin{figure}[h!]   
\centering 
\includegraphics[width=7.5cm, height=5.5cm]{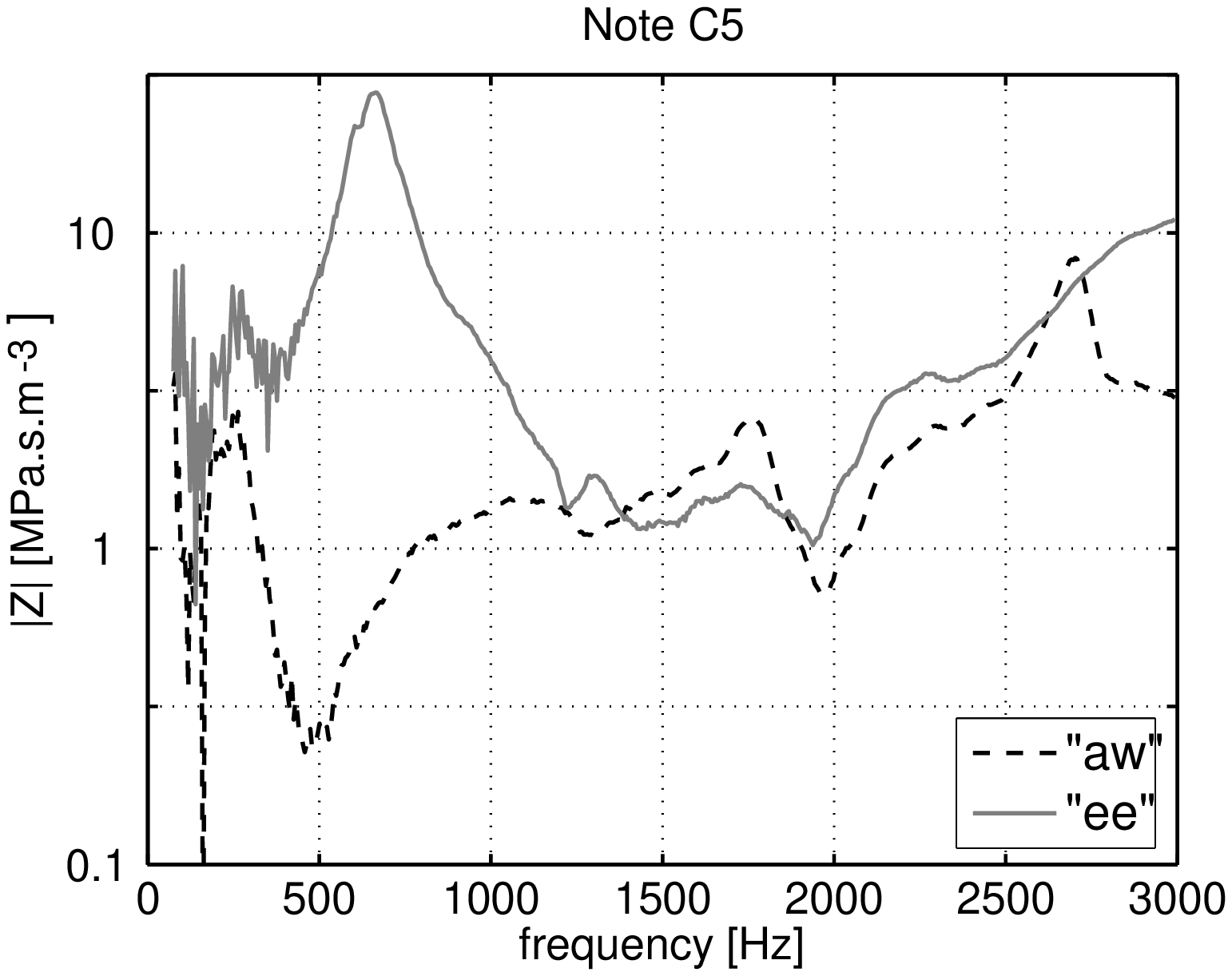} 
\includegraphics[width=7.5cm, height=5.5cm]{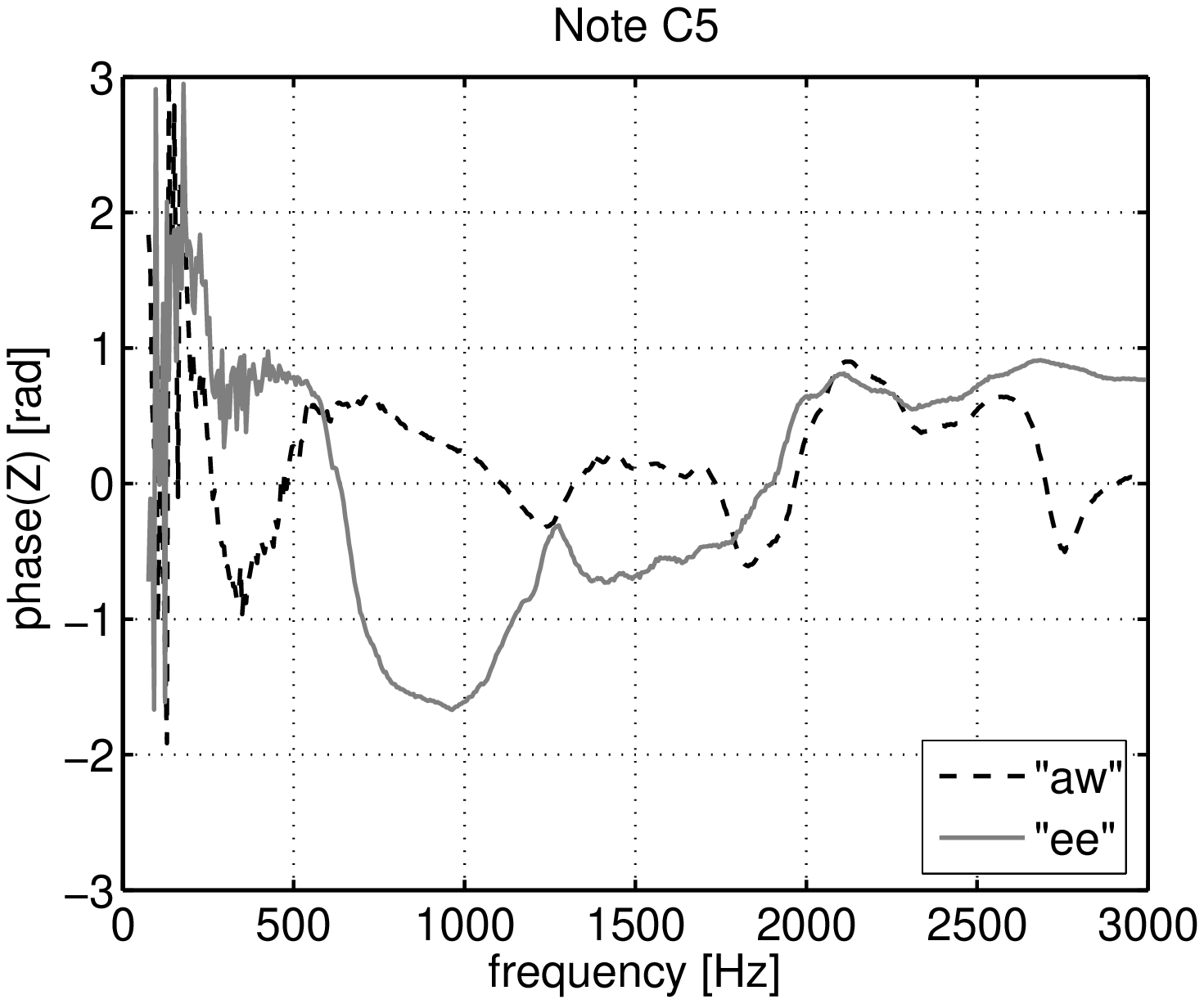}   
\caption{The impedance spectra of player D's airway for two
configurations described as ``ee'' and ``aw'', for the note C5.}   
\label{fig:DiffIntra1} 
\end{figure}

For several players, the ``ee'' configuration produced a strong peak
between 560 and 1000 Hz, a peak that is associated with the
constriction between tongue and palate (see FIG.~\ref{fig:DiffIntra1}). For
many players, however, the configuration they produced when asked to
mime the ``ee'' embouchure, had no such peak and indeed resembled
somewhat the impedance measured when they were asked to mime the ``aw''
embouchure. However, the average level of impedance, even for these
players, was in general higher for ``ee'' than for ``aw''. Not all
players use the ``ee'' and ``aw'' terminology for the embouchure and it is
possible that the instruction was in this case confusing. It is
important to remark that this terminology in terms of vowels refers
more to the position of the tongue in the mouth than to the real
configuration of the vocal tract in speech as the mouth of the player
is of course closed around the mouthpiece.

\subsection{Differences among players for ``normal'' playing mode} 

\begin{figure}[h!]   
\centering 
\includegraphics[width=7.5cm, height=5.5cm]{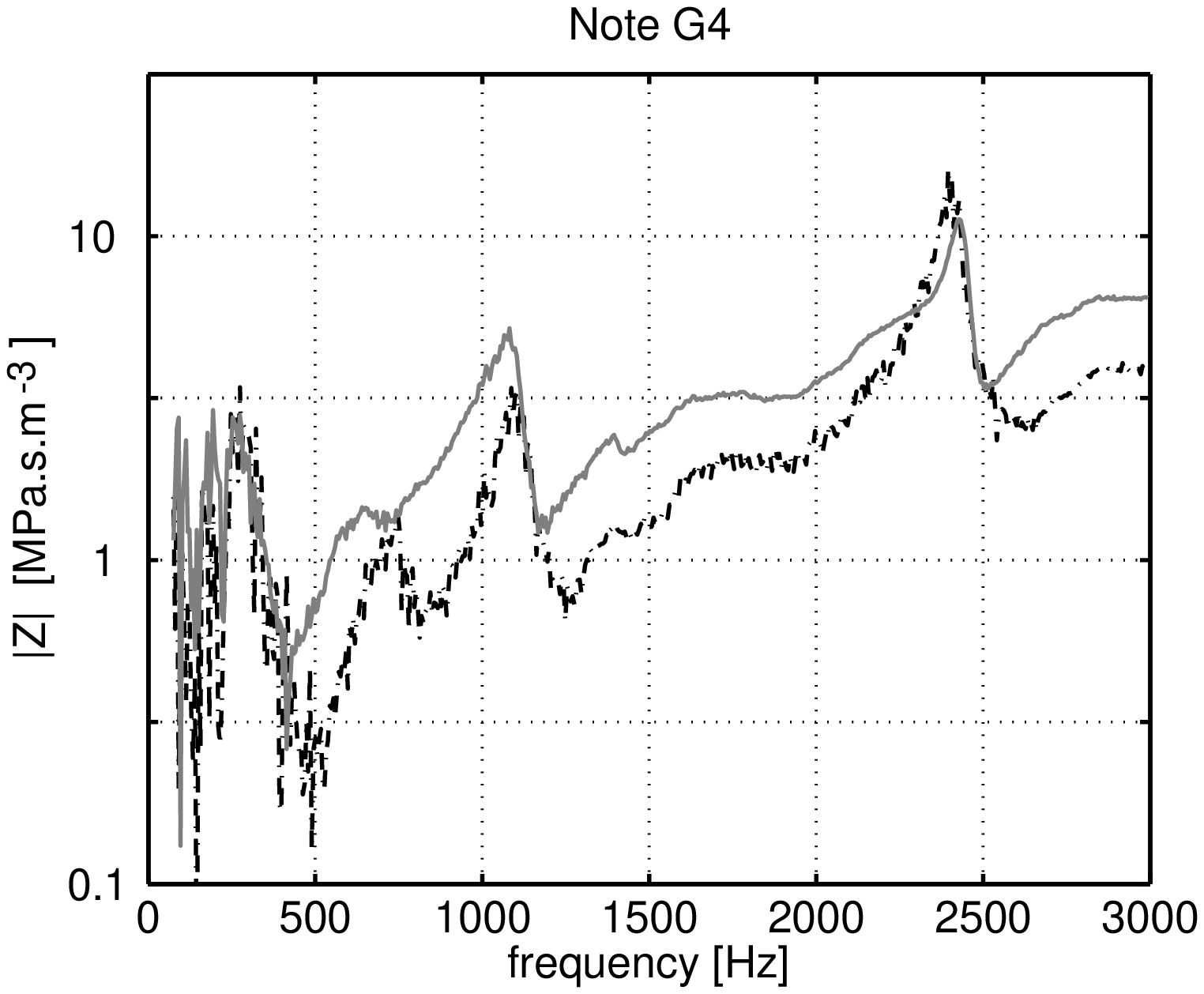} 
\includegraphics[width=7.5cm, height=5.5cm]{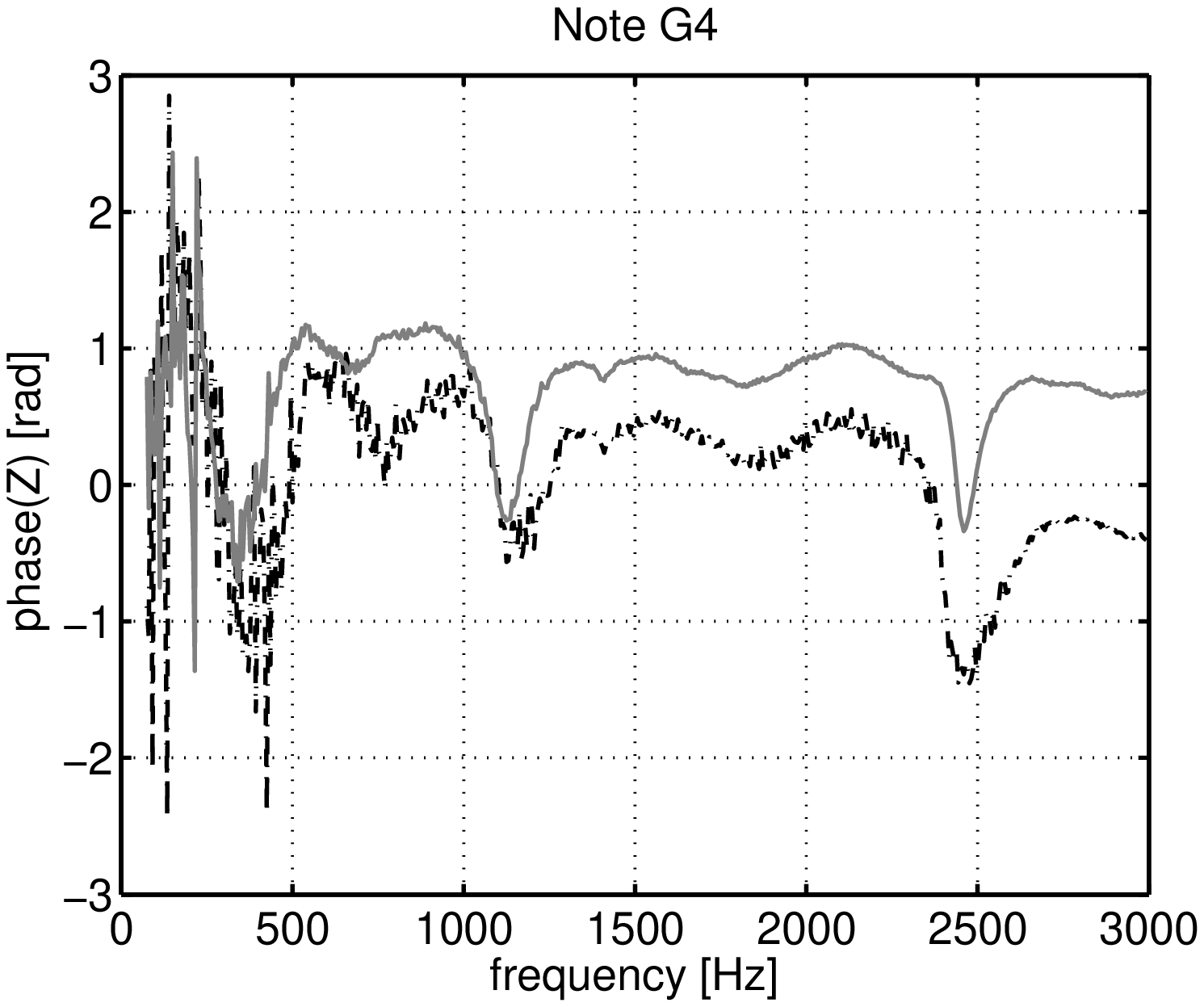} 
\includegraphics[width=7.5cm, height=5.5cm]{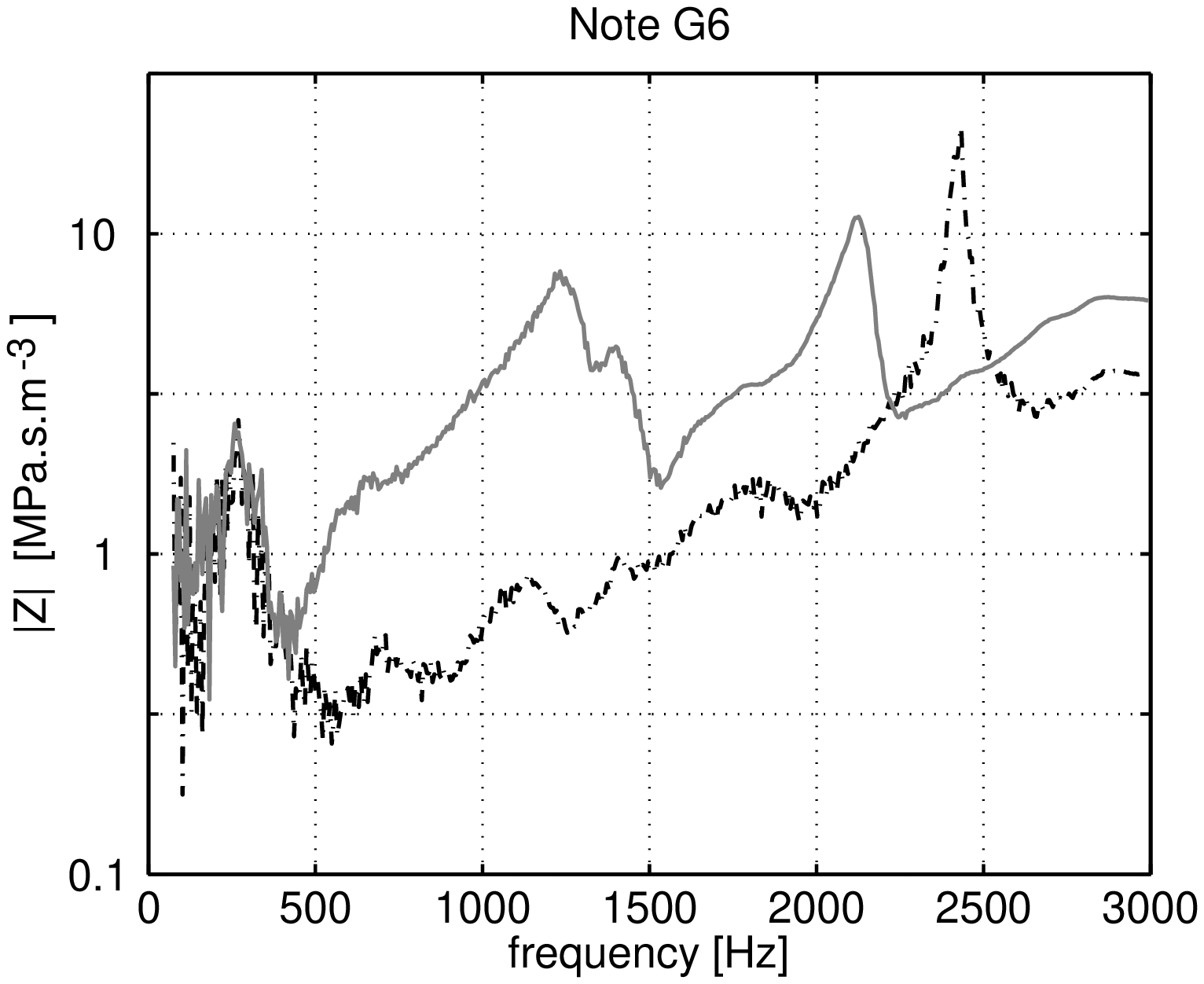} 
\includegraphics[width=7.5cm, height=5.5cm]{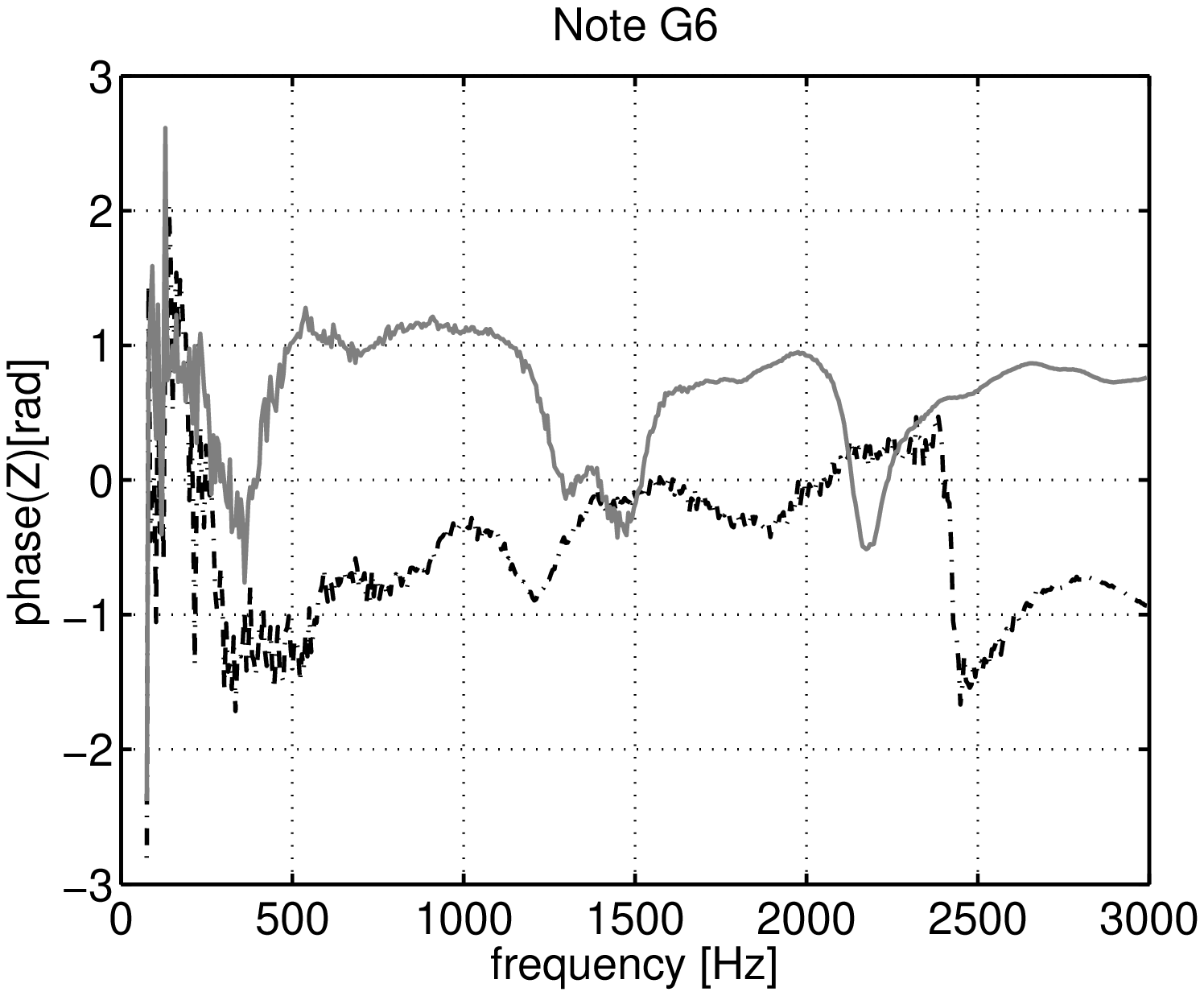} 
\caption{The impedance spectra of the respiratory airway of two
experienced professional musicians (player B in black and player E in
grey), for notes G4 (top) and G6 (bottom).}    
\label{fig:DiffInter} 
\end{figure}

 We study here the configurations that musicians use in ``normal''
 playing, which means the configuration they adopt usually, when they
 have no special musical intentions, in the \emph{mezzo forte}
 nuance. For comparisons, we choose the note G4 which is
 representative of the low and medium register and the note G6 for the
 high register. In FIG.~\ref{fig:DiffInter}, the same two musicians
 mime playing each of the notes.

The configuration for the note G4 is qualitatively similar for both
musicians. A few exceptions apart, it is a configuration used by many
players in the normal playing mode for almost the whole range of the
clarinet, as shown in the figures available in~\cite{webVT}. However,
the configuration adopted for the very high register can differ quite
considerably among players: some musicians adopt a configuration that
enhances the second peak and moves it into the frequency range of
the note played whereas some others tend to adopt a configuration that
reduces the amplitude of this peak.

\subsection{Variations used by players}  

\begin{figure}[h!]   
\centering   
\includegraphics[width=7cm, height=5.5cm]{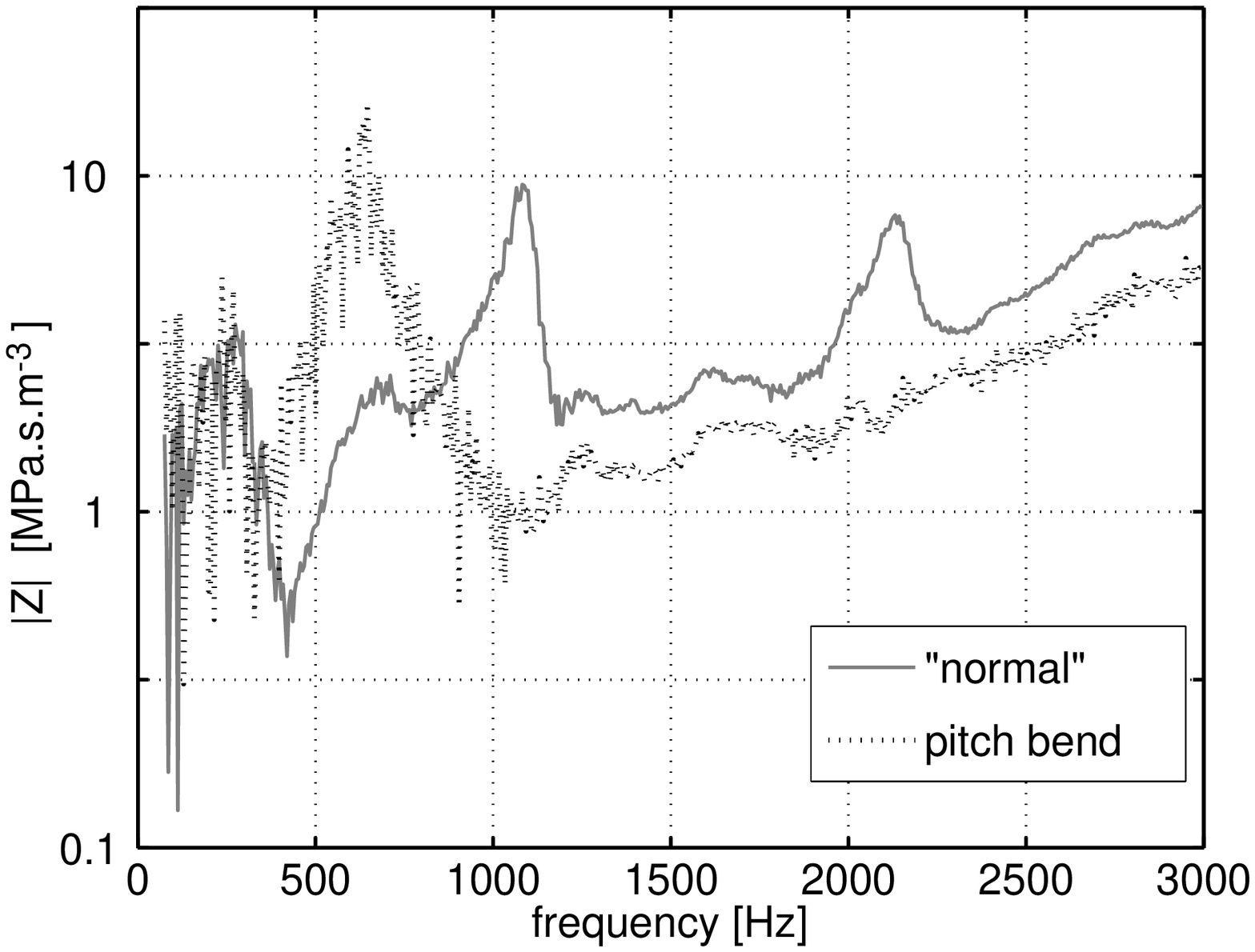} 
\includegraphics[width=7cm, height=5.5cm]{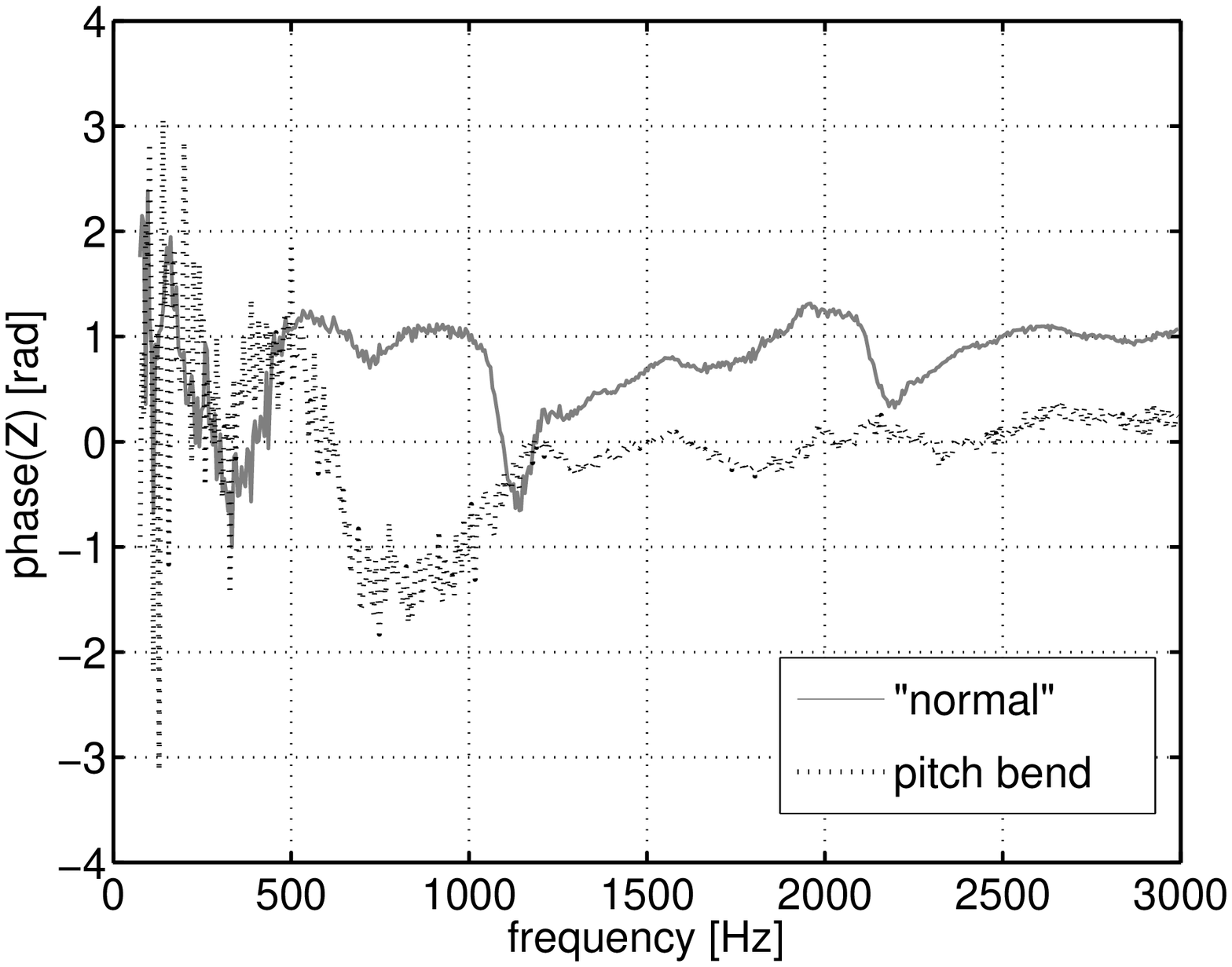} 
\caption{A comparison of the impedance spectra measured on player C for
the configurations for normal playing and for performing a pitch bend.}   \label{fig:normal-bend} 
\end{figure}

Players agree that they use different embouchures for different
effects.  The embouchure includes the lip and jaw position, and hence
the force, the damping and the position on the reed may vary. The
aspect being studied here is the way in which the mouth or vocal tract
geometry changes can affect the impedance spectrum. The substantial
changes shown in FIG.~\ref{fig:normal-bend} suggest that this latter
effect may not be negligible. even if the configuration in normal playing
is quite stable over the whole register.

\begin{figure}[h!]   
\centering   
\includegraphics[width=7cm, height=5.5cm]{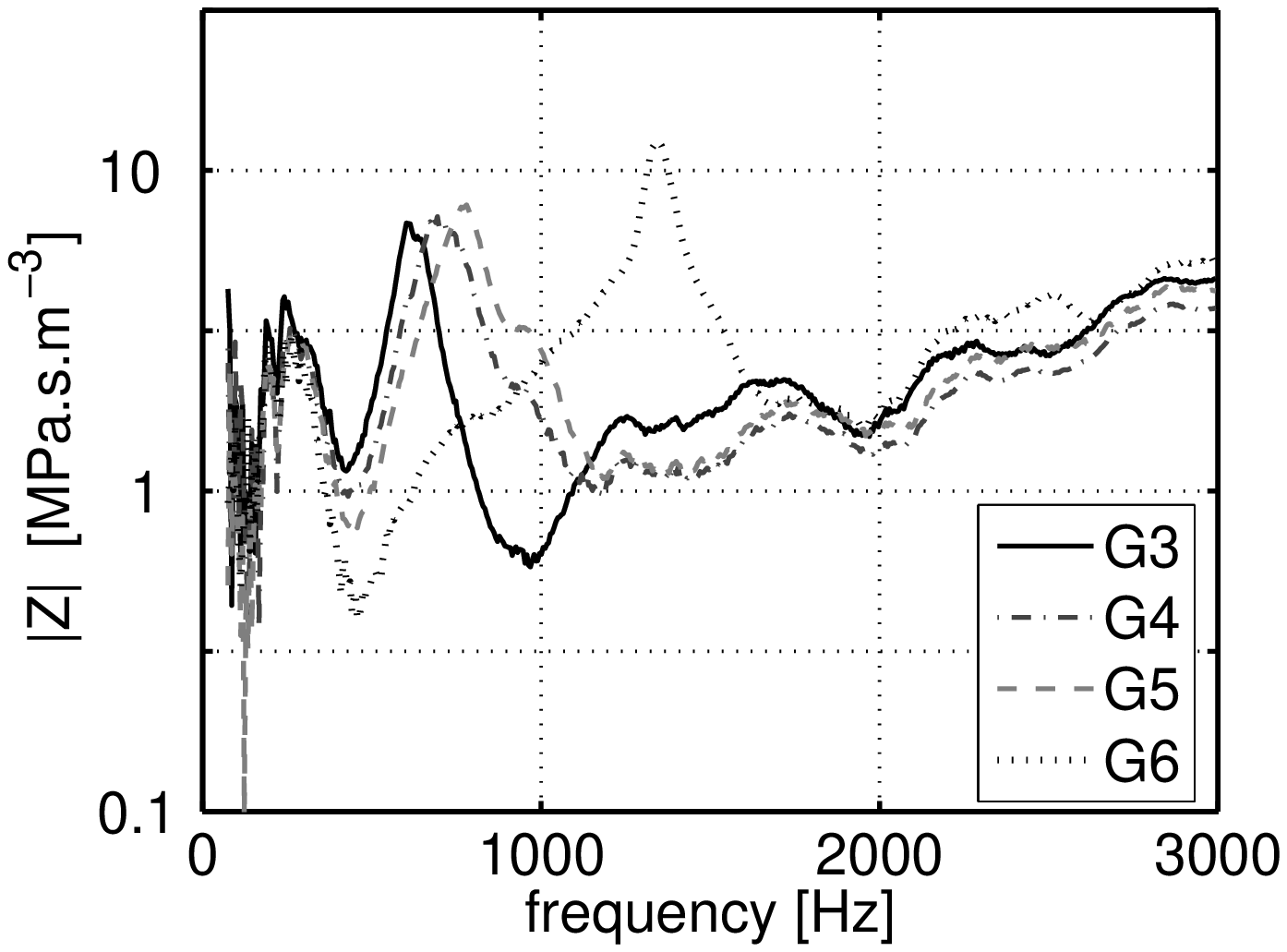} 
\includegraphics[width=7cm,height=5.5cm]{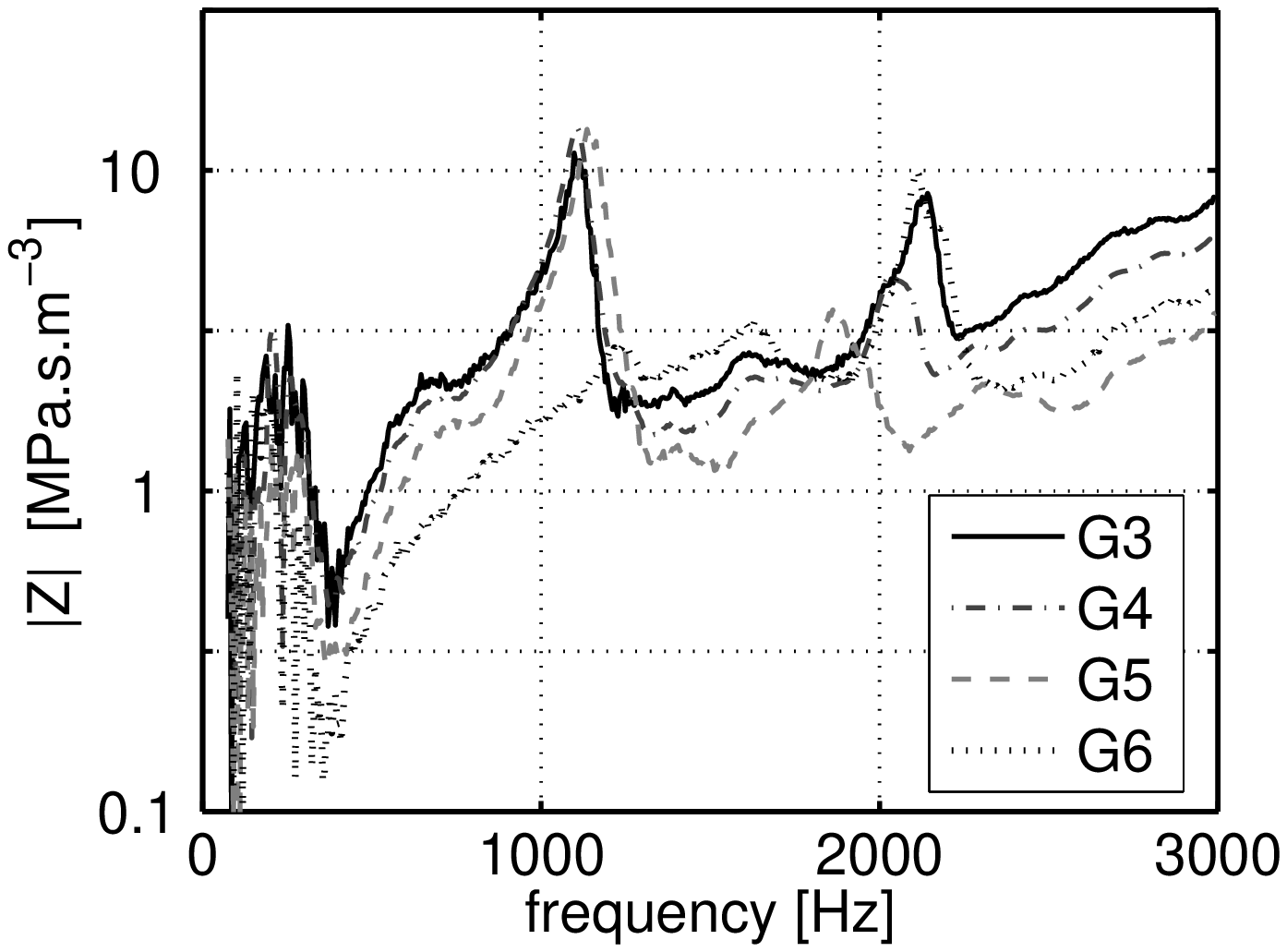} 
\includegraphics[width=7cm,height=5.5cm]{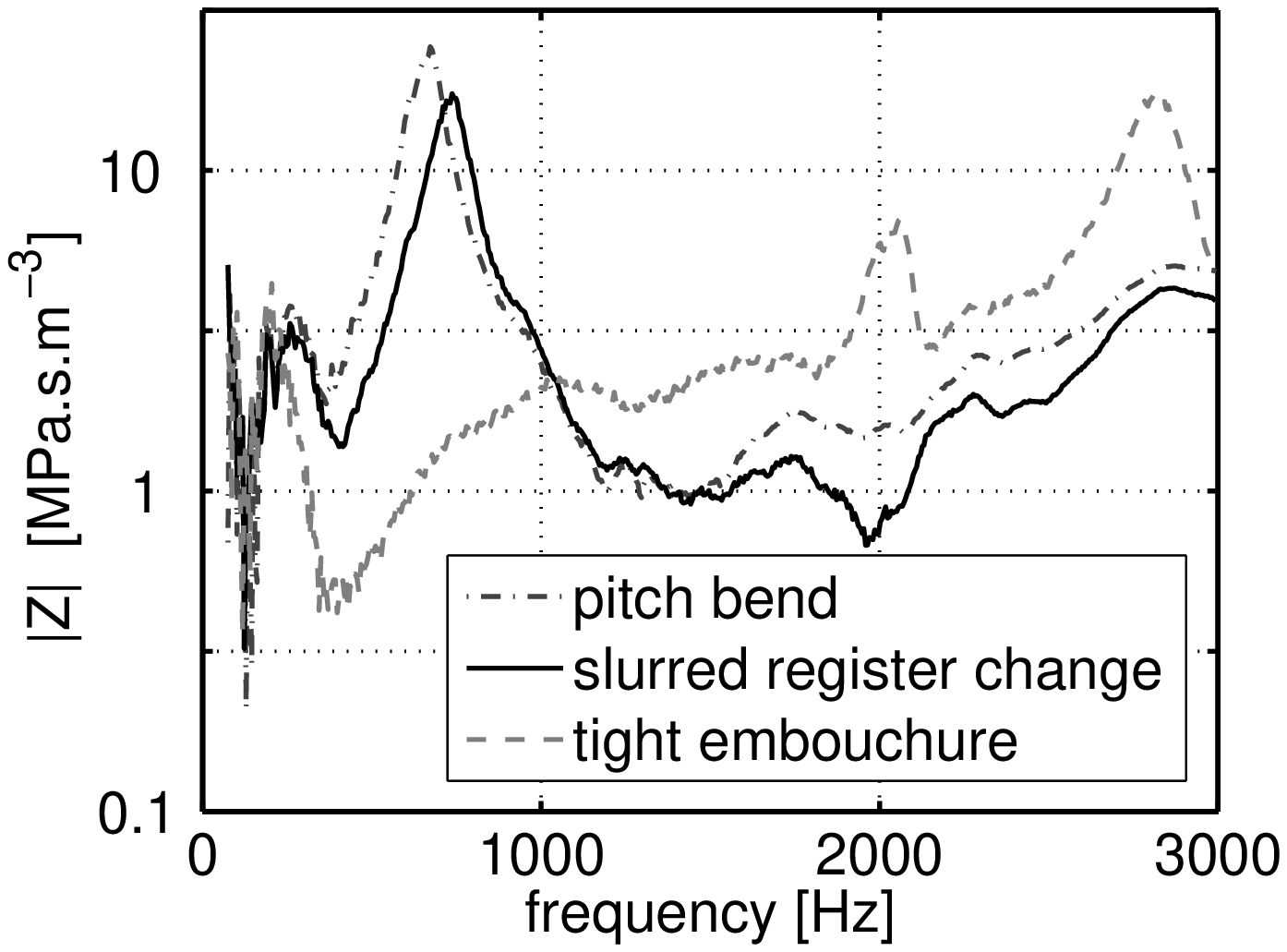} 
\includegraphics[width=7cm, height=5.5cm]{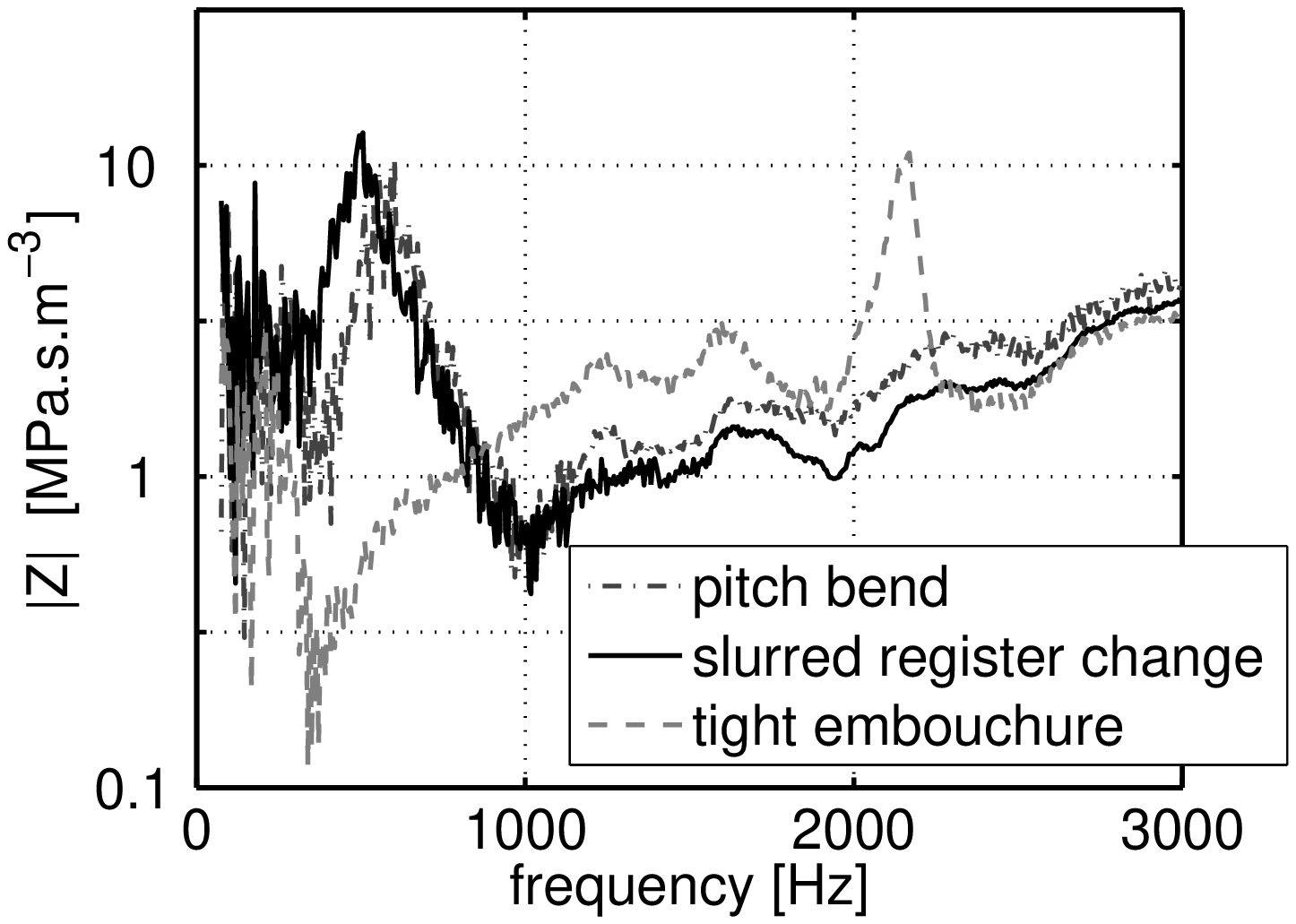}   
\caption{A comparison between two professional players (player D at left
and C at right) for their normal playing configurations
  (top) and for those used for some less usual effects (bottom).}
\label{fig:MargLawr} 
\end{figure}

 It is
interesting to note the remarkable similarity in the impedances for
``special effects'' between two professional players who played
together for several years in a major national orchestra, whereas they
do not adopt the same configuration for normal playing
(FIG.~\ref{fig:MargLawr}).

One of the professional players expressed her control of pitch and
timbre thus: she uses a ``ee''  for 
the high register or for brightening the sound and in contrarily she
adopts a ``aw'' configuration for darkening the timbre and lowering
the pitch. The differences between these two configurations are
represented in FIG.~\ref{fig:DiffIntra1}.


\subsection{Subtlety}  

\begin{figure}[h!]   
\centering  
\includegraphics[width=7.5cm, height=5.5cm]{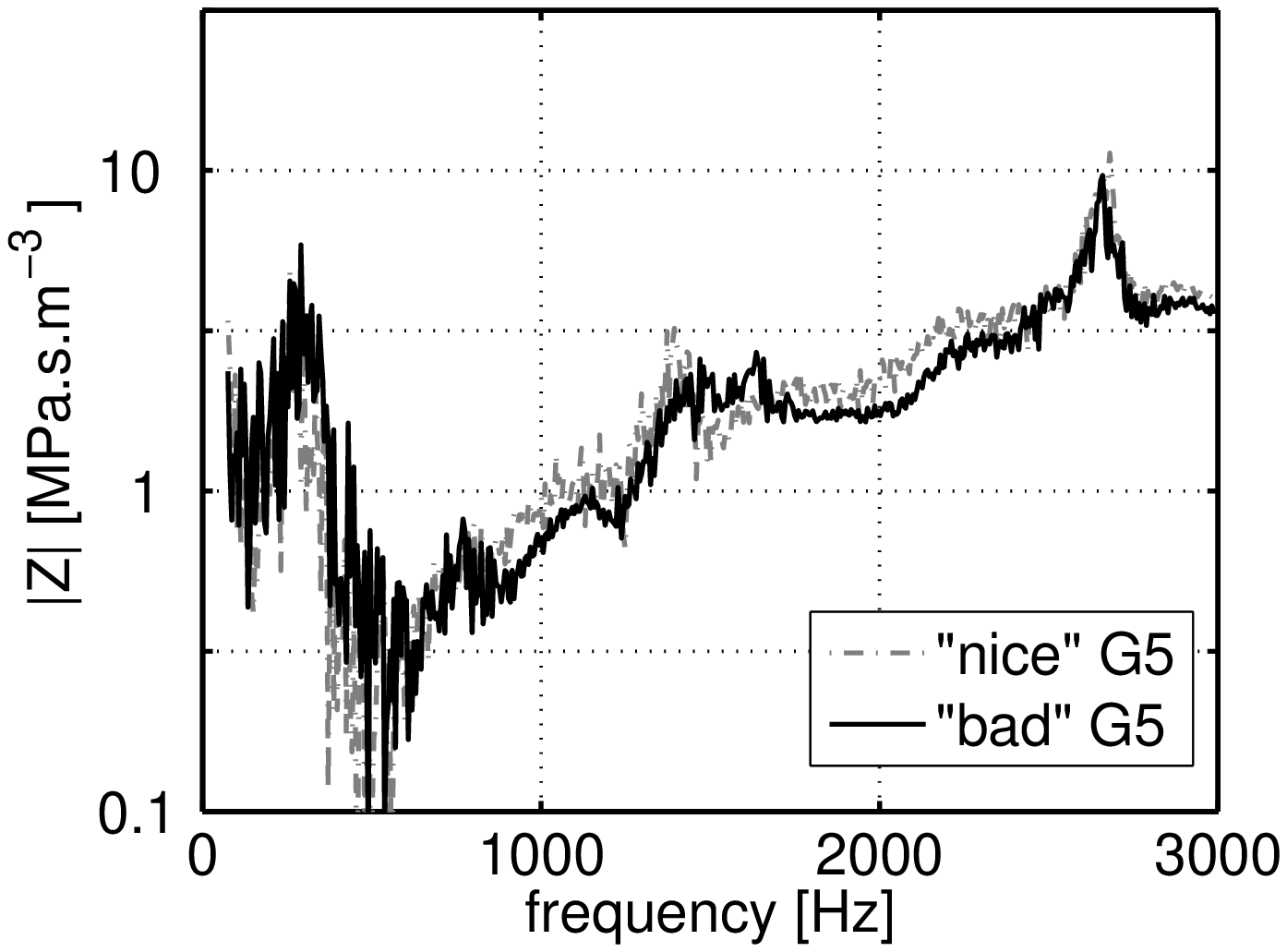} 
\includegraphics[width=7.5cm, height=5.5cm]{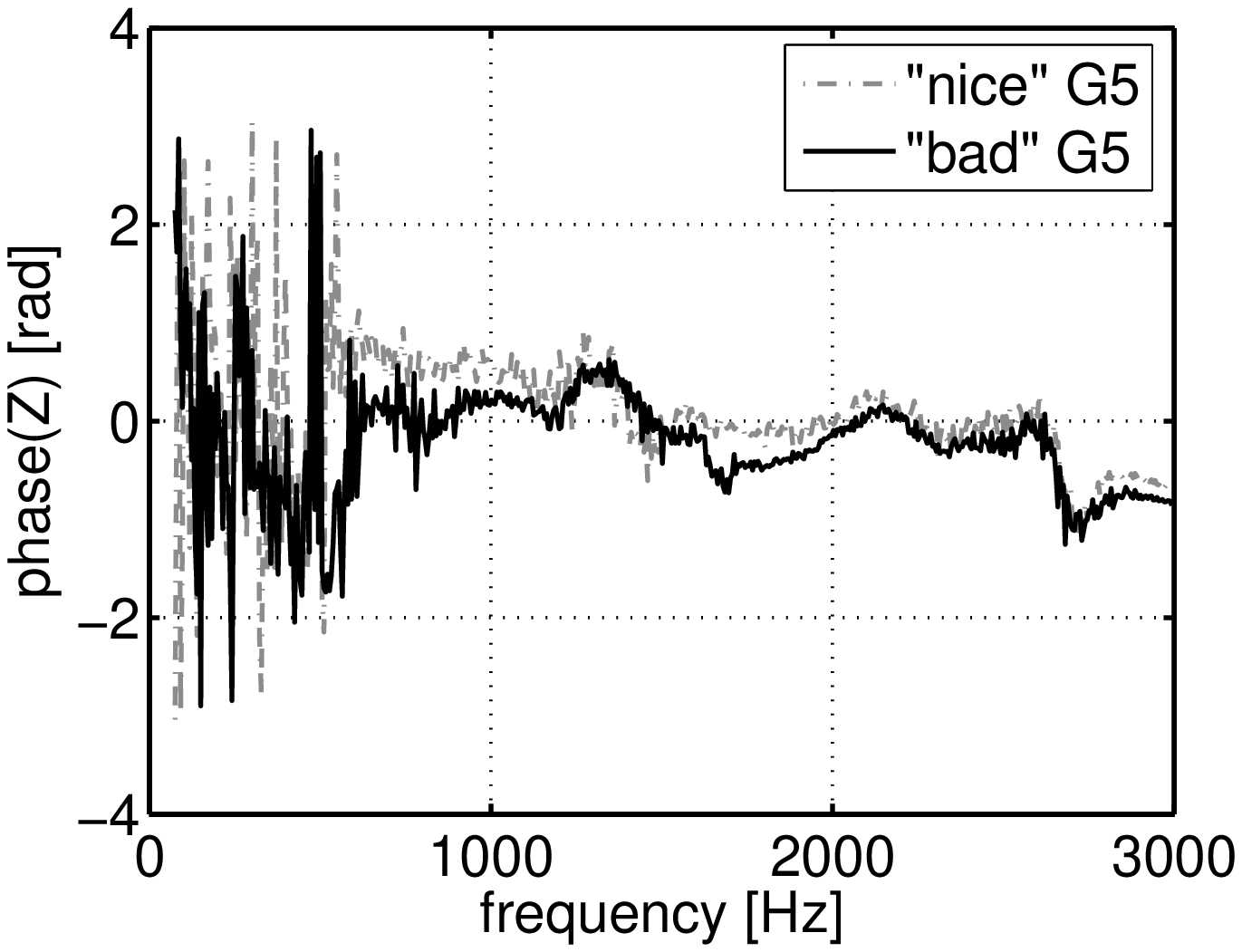} 
\caption{The impedance spectra of the vocal tract of player A for two
configurations associated with a ``nice'' sound or with a ``bad'' one,
i.e. one which is to be avoided (note G5).}   \label{fig:cathmc_nicebad} 
\end{figure}

In most cases, different tract configurations that were reported to be
used to produce different effects on the sound were found to have
different impedance 
spectra. However, for some of the players, the
impedances measured when they were miming ``good'' and ``bad'' embouchures
differed by amounts comparable with the measured reproducibility of a
single embouchure. For example, FIG.~\ref{fig:cathmc_nicebad} shows
a large similarity between the impedances for embouchures described by
a very experienced soloist as those corresponding to a ``nice'' and a
``bad'' sound. We presume that in this case the differences had more to
do with aspects of the embouchure such as lip tension and position,
and less to do with the tract configuration.

\subsection{Summary of the measurements}
The players are classified into groups for which the impedances look
similar for notes G3 to G5. The results of one player contained
unexpected and unexplained high levels of noise, and are
omitted. Results are summarised in TABLE~\ref{tab:resume}.

\begin{table}[h!]
\enlargethispage{0.5cm}
  \centering
\rotatebox{90}{
\begin{tabular}{|p{3.2cm}|p{4.9cm}|p{4.9cm}|p{4.9cm}|}
\hline
\centerline{players} & \multicolumn{2}{c|}{normal playing} & \centerline{special effects}\\
\cline{2-4}
 & \centerline{G3, G4, G5} & \centerline{G6} & \centerline{pitch bend} \\
\hline
A, B, C, E and one other & 2 resonances: 1000-1400 Hz (3.2
$10^6$-$10^7$ Pa.s.m$^{-3}$) and
2100-2400 Hz. & The second resonance disappears for players A, C and
D. For player F, the second resonance is lowered by 500 Hz. & Only one resonance: 600 Hz (1.3 $10^7$
Pa.s.m$^{-3}$ ) for player D, 1400 
Hz (4 $10^7$ Pa.s.m$^{-3}$) for F, 2500 Hz for A and 2700 Hz (in the
last two cases, it 
is actually the first resonance which disappears). \\
\hline
Player D & One resonance: 700 Hz, 5.6 $10^6$ Pa.s.m$^{-3}$. & This
resonance is shifted to 1400 Hz. & Strong resonance (1.8 $10^7$
Pa.s.m$^{-3}$) at 700 Hz 
which suggests that she uses a ``ee'' configuration. \\
\hline
Player H, F and 4 other players & The impedance grows continuously with
the frequency, with a slope of 30 dB for 2500 Hz. 2 small resonances
at 1200 Hz and 2000-2300 Hz. & The second resonance is slightly
enhanced. &  For half of the players, the resonances disappear; for
the others (like player E), the first resonance is strongly enhanced.\\
\hline
Player G & G3-G4: 2 acute resonances at 1200 and 2100 Hz. G5: one strong
resonance at 1400 Hz. & As for G5. & As for G3 and G4. \\
\hline
2 players & \multicolumn{3}{p{14.7cm}|}{Same configuration for all playing
modes. 1 strong resonance at 2200 or 2500 Hz, between 3.2 $10^7$ and $10^8$ Pa.s.m$^{-3}$.}\\
\hline
\end{tabular}}

  \caption{Summary of the features of impedance spectra measured on 16
    players for the vocal tract configurations they used for the
    different cases listed.}\label{tab:resume}
  
\end{table}

\section{Theoretical model}

\subsection{Model for the vocal tract}\label{sec:ModCV}

This model draws on work in speech science, where scientists are more
interested in what happens at 
the glottis and usually calculate either the transfer function or the
impedance at the glottis. However, numerical simulations which were
done in speech synthesis to calculate the impedance at the glottis can
be used in our study by inverting the calculation and using the
appropriate impedance at the glottis. The model used is the one
developed by Sondhi~\cite{Son74,Son87} with yielding walls. The vocal
tract is represented by concatenated cylinders and the relation
between the variables at the input of cylinder $k+1$ and cylinder $k$
(k=0 at the glottis) is the following:

\[
\begin{pmatrix}
  p_{k+1} \\
u_{k+1}
\end{pmatrix}
= 
\begin{pmatrix}
  A & B \\
  C & D
\end{pmatrix}
\begin{pmatrix}
  p_{k} \\
u_{k}
\end{pmatrix}
\]

with
\begin{eqnarray}
  \label{eq:1}
  A&=&D=\cosh(\frac{\sigma l}{c}) \nonumber\\
 B&=&\frac{\rho c}{S}\gamma \sinh(\frac{\sigma  l}{c}) \quad C=\frac{S}{\rho c}\gamma \sinh(\frac{\sigma  l}{c})
\end{eqnarray}
with
\begin{eqnarray}
  \label{eq:2}
 \gamma& =& \sqrt{\frac{r+j\omega}{\beta + j\omega}}\\
\sigma& =& \gamma(\beta + j\omega)\\
\beta &=&\frac{j\omega \omega_t^2}{(j\omega+r) j\omega +\omega_w^2 }+\alpha\\
\alpha&=&\sqrt{j\omega q}
\end{eqnarray}
where $r$ and $\omega_w$ are related to the yielding properties of the
vocal tract and represent respectively the ratio of wall resistance to
mass and the mechanical resonance frequency of the wall. Their values
are set to $r=408$rad.s$^{-1}$ and $\omega_w/2\pi=15$Hz. $\omega_t$ is the frequency of a resonance of the tract when sealed at both the
glottis and the lips, and which is associated with the finite mechanical
compliance of the walls (like the ``breathing mode'' in a string
instrument): $\omega_t/2\pi=200$Hz. The parameter
q is a correction for thermal conductivity and viscosity, which is set
to $q=4$rad.s$^{-1}$.

The calculation was done was done using a program written by Story~\cite{Sto00}, except that it was inverted in order to
calculate the impedance at the mouth (and not at the glottis).
To complete the calculation we need the glottis geometry (tube 0) as,
in contrast with speech, the vocal folds are not entirely closed. We also
need the boundary condition at the glottis: $p_0=Z_{sg}u_0$, where
$Z_{sg}$ is the input impedance of the subglottal tract.

\subsection{The glottis}

According to the laryngoscopic study by Mukai~\cite{Muk92}, the glottis
of professional wind musicians is usually a narrow slit, to which our
first order approximation is a rectangle of length $a=10$ mm, width
$b=1.5$ mm and thickness $e=3$ mm.

However, the discontinuity between the cross section of this slit and
that of the trachea requires an acoustic end correction. Here the
glottis is treated as a tube baffled at both ends, having an effective
length of $e_{eff}=e+0.85r_g$ where $r_g$ is the equivalent radius
\footnote{In principle, the end correction for a slit is greater than 
that for a circular aperture of the same area \cite{Mor00}. However,
this approximation is appropriate, given the experimental
uncertainties.} of the glottis ($r_g=\sqrt{ab/\pi}=2.2$ mm).

Consequently, cylinder 0, which represents the glottis and thus connects
the vocal tract to the subglottal tract in the model, has a length
$e_{eff}=4.7$ mm and area $ab$.

\subsection{The subglottal tract}

\begin{figure}[h!]
  \centering
  \includegraphics[height=2.3cm, width=11.5cm]{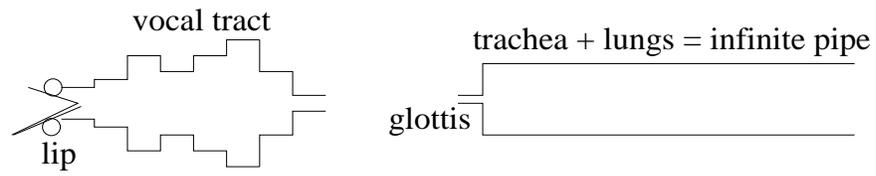}
  \caption{A schematic of the waveguide used to model the respiratory
airway. It is not to scale, and the number of elements has been reduced
for clarity. The clarinet mouthpiece is inserted between the lips at left.}
  \label{fig:airway}
\end{figure}

For all but very low frequencies, the results depend only very weakly on
the subglottal tract, so we use a very simple model. (FIG.~\ref{fig:airway})


The lungs are very lossy at the frequencies of interest, and so
reflections are minimal. For that reason, they are modelled here as a
purely resistive load with the same characteristic impedance (i.e. an 
infinitely long pipe whose radius equals that of the trachea, $r_t=$9
mm) 
\begin{equation}
 Z_{sg}=\frac{\rho c}{ \pi r_t2} 
\label{eq:Rpoumons}
\end{equation}

\subsection{Correction at the input of the mouth}

In comparison with the cross sectional area of the mouth, the area of
the reed inside the mouth is small, as is that of the impedance head. It
resembles thus a small piston vibrating in a baffle that seals a larger
waveguide, or a discontinuity in waveguides, which is often modelled by
adding an end correction to the smaller element. Physically, the volume
of air in the end correction has an inertance comparable to the that of
air in the strongly diverging part of the radiation field in the larger
guide. The end correction for a baffled pipe is used at this end of the
vocal tract, too: an element with radius $r=$3.9 mm and length  $l=0.85r$
\cite{Fle95}.

As we used mainly the MRI data from Story and Titze~ \cite{Sto96}, we
divided the vocal tract, of length 170.4 mm, in finite elements of
length 4 mm, giving 44 elements, plus the zeroth element representing the  
glottis (this one is actually divided in two elements: the first
has the same length as the others whereas the second is used to
adjust the effective length of the glottis)\label{tubes}. A complication is due to
the insertion of the mouthpiece about 
10 mm in the mouth, which puts the first two tract elements effectively
in parallel with the rest. The iterative calculation of section~\ref{sec:ModCV}
is conducted on 33 tract elements, beginning at the glottis, which is
loaded with the subglottal resistance. This gives the impedance $Z_1$.
The two elements closest to the mouth and sealed at the mouth end give
an impedance $Z_2$. The total uncorrected impedance is therefore

\begin{equation}
  \label{eq:ZCVnc}
  Z_{nc}=\frac{Z_1Z_2}{Z_1+Z_2}
\end{equation}
which, when the end correction mentioned above is added, gives:
\begin{equation}
  \label{eq:ZCV}
  Z_{CV}=Z_{nc}+jZ_0\frac{\omega}{c}l
\end{equation}
with $Z_0=\rho c/(\pi r2)$.


The result of end effects and
the parallel elements at the mouth are noticeable primarily at high
frequency ($>$ 3000 Hz). However, at such frequencies the unidimensional
model fails anyway because of its neglect of transverse modes. For
example, El-Masri \emph{et al.}~\cite{ElMas98} showed how the plane
wave approximation gave poor predictions for the behaviour of the tract
at frequencies above 4500 Hz. Further refinement is therefore
inappropriate in this simple model. What is of practical importance in
these corrections is their substantial reduction of the amplitude of
the resonances at high frequency, which avoids artifacts in simulations that
have strong high frequency resonances that do not correspond to those of
the tract made of flesh. Further, the cutoff frequency of an array of
open tone holes in the clarinet 
is typically around 2000 Hz, so our interest need not exceed the range
0-3000 Hz.

\subsection{Results for two vowels. Adjustment.}

The aim of this simulation is to 'invert' the model, i.e. to obtain the
area function from the impedance spectrum. Solutions to inversion are
not unique, but other constraints on vocal tract shape eliminate many.
An inversion was obtained with assistance from Brad Story, which mapped
the calculated impedance of the complete model, including corrections to
the area function. The program begins with the first three resonance
frequencies and the mapping was generated using area functions such as
given by \cite{Sto96} (using 4 mm elements).  In general it was found
that the impedance spectra calculated from the mapped area functions
differed noticeably from the original impedance measurements. A further
program was therefore written to allow iteration by local adjustment of
the area function to improve the fit. By such iterations, anatomically
possible area functions giving the experimental impedance spectra were obtained.

Two important results are shown. Clarinettists often refer to two tract
configurations as ``ee" and ``aw", being those for the vowel sounds
(those of ``heed" 
and ``hoard") sensibly resembled by the playing position. The resemblance
is only approximate, of course: for the vowels, the mouth is open to
different extents, whereas in the playing configuration the mouth is
sealed by the mouthpiece of nearly constant area.

\begin{figure}[h!]
     \centering
     \includegraphics[height=5.5cm, width=7cm]{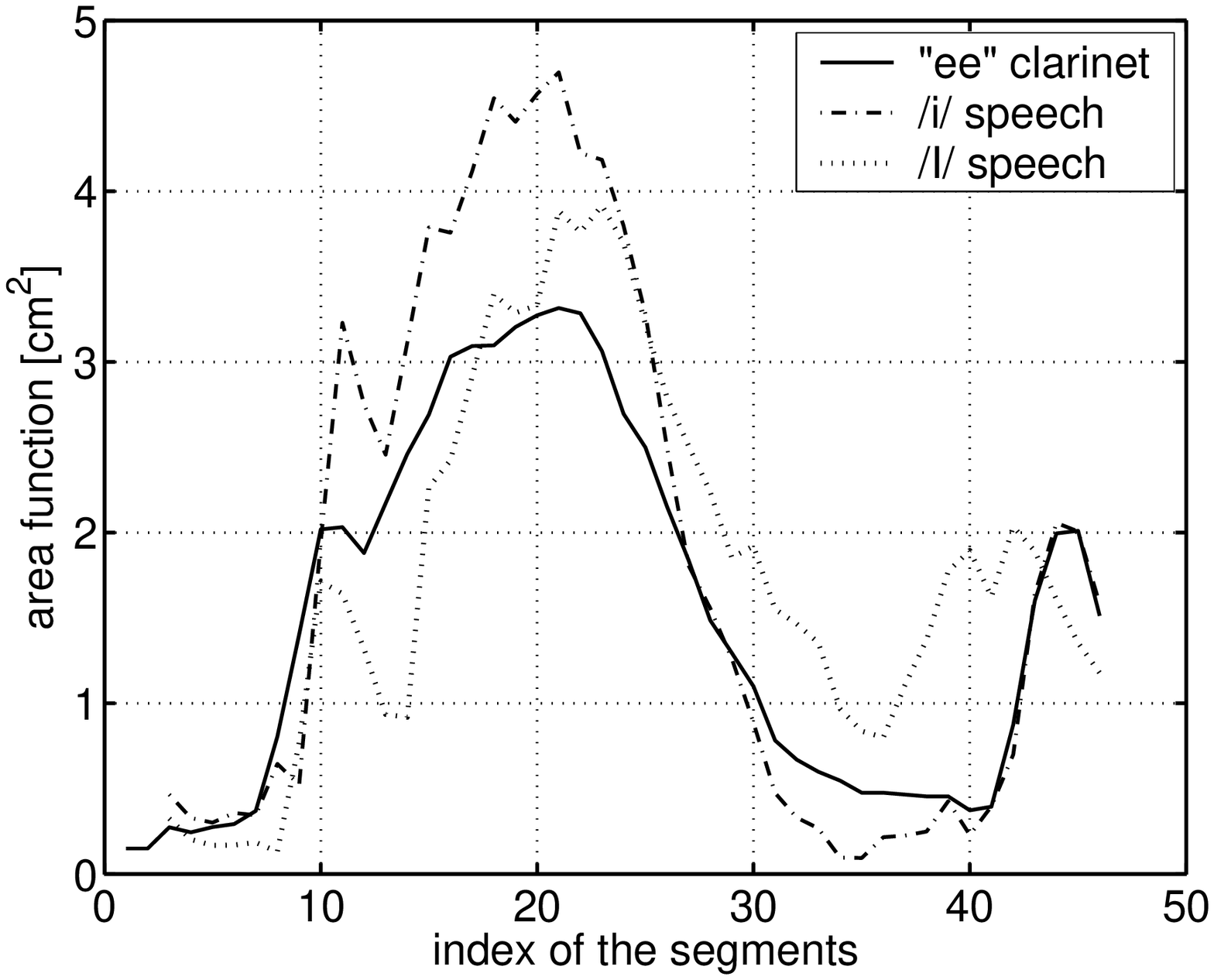}
  \includegraphics[height=5.5cm, width=7cm]{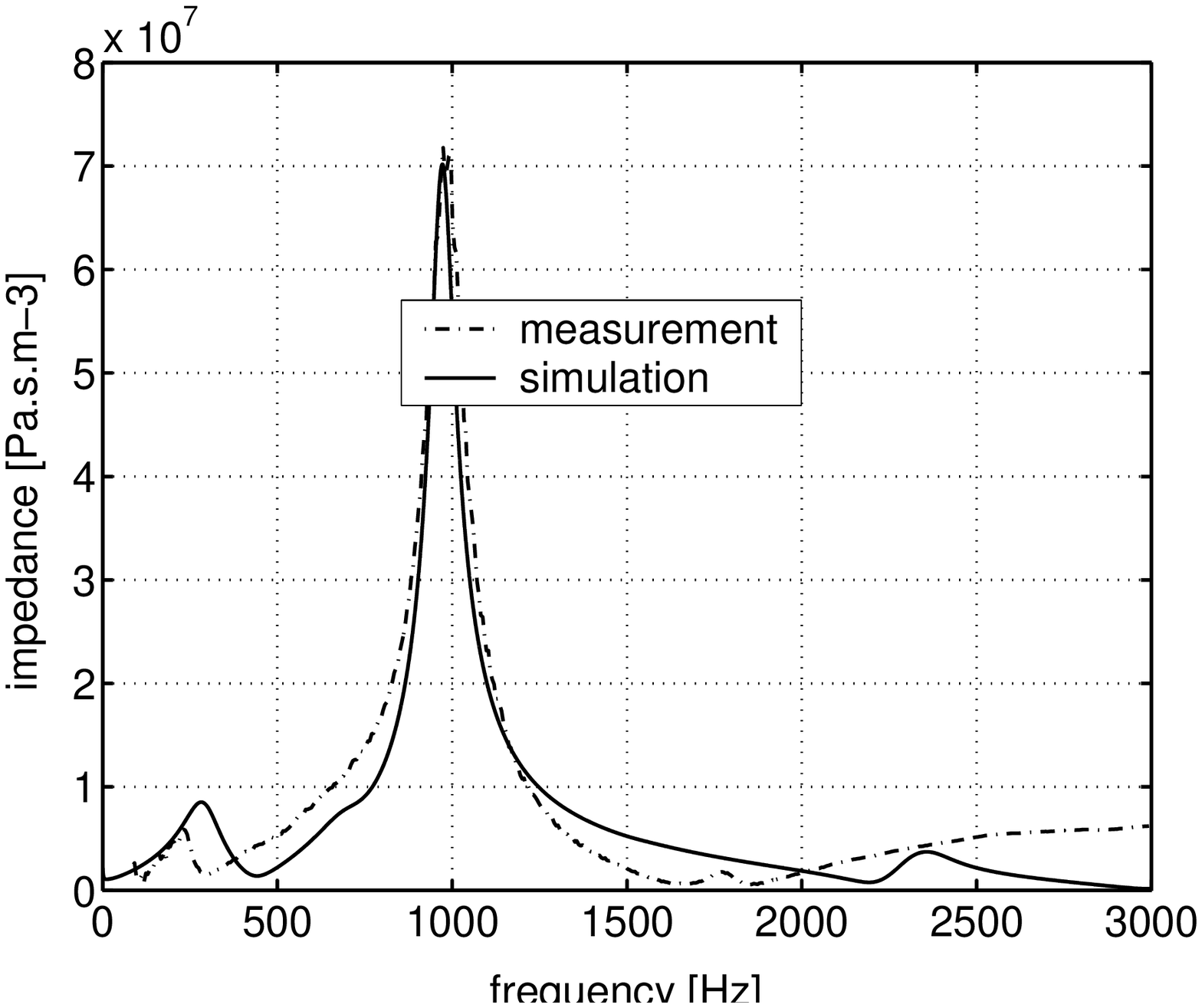}
  \caption{At left: area function for the vowels /i/ and
    /\textsci/ (data from Story and Titze with glottis added). Also
shown are the calculated tract configuration named ``ee" by
clarinettists, as determined from inversion. The abscissa is the element
number: the first two are the glottis (see~\ref{tubes}), the next 44,
of length 4 mm 
comprise a tract 174 mm long.  At right: the impedance spectrum for
the ``ee" tract configuration measured on one of the authors (player G) an
amateur clarinettist.}
    \label{fig:ee}
   \end{figure}

For ``ee", FIG.~\ref{fig:ee} shows the plausibility of the area
function, which is rather similar to that of the vowels /i/ (``heed'')and
/\textsci/ (``hid''). This configuration has a cross section at the palatal
constriction lying between the values reported for the two vowels, which in
English differ little except in their duration. At the mouth end, the
area function is set equal to the cross section of the clarinet
mouthpiece, 15 mm from the end.\\ 
It proved impossible, however, to fit the first peak, for which the
inversion gave frequencies that were systematically too
high  (e.g. between   230 and 250 Hz instead of 200 Hz for ``aw" and
between 250 and  280 Hz instead of 230 Hz for ``ee''). 

A resistance of 1.5 MPa.s.m-3 was added at the glottis. Although this
was an empirical adjustment to fit the measured height of the peak
values of impedance, it might be justified by considering that energy
would be lost from propagating waves due to turbulence produced by
flow through this narrow slit. 

\begin{figure}[h!]
     \centering
     \includegraphics[height=5.5cm, width=7cm]{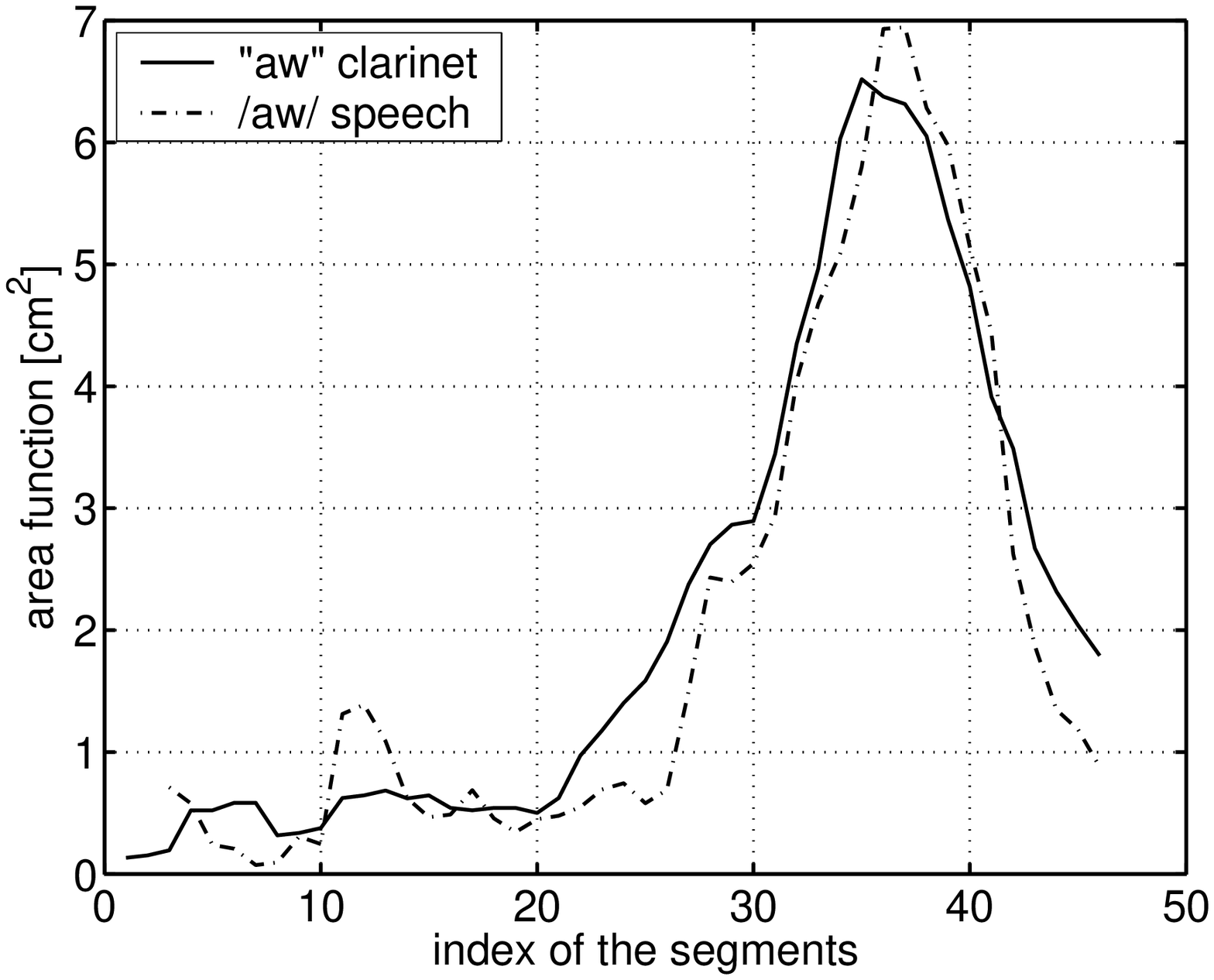}
  \includegraphics[height=5.5cm, width=7cm]{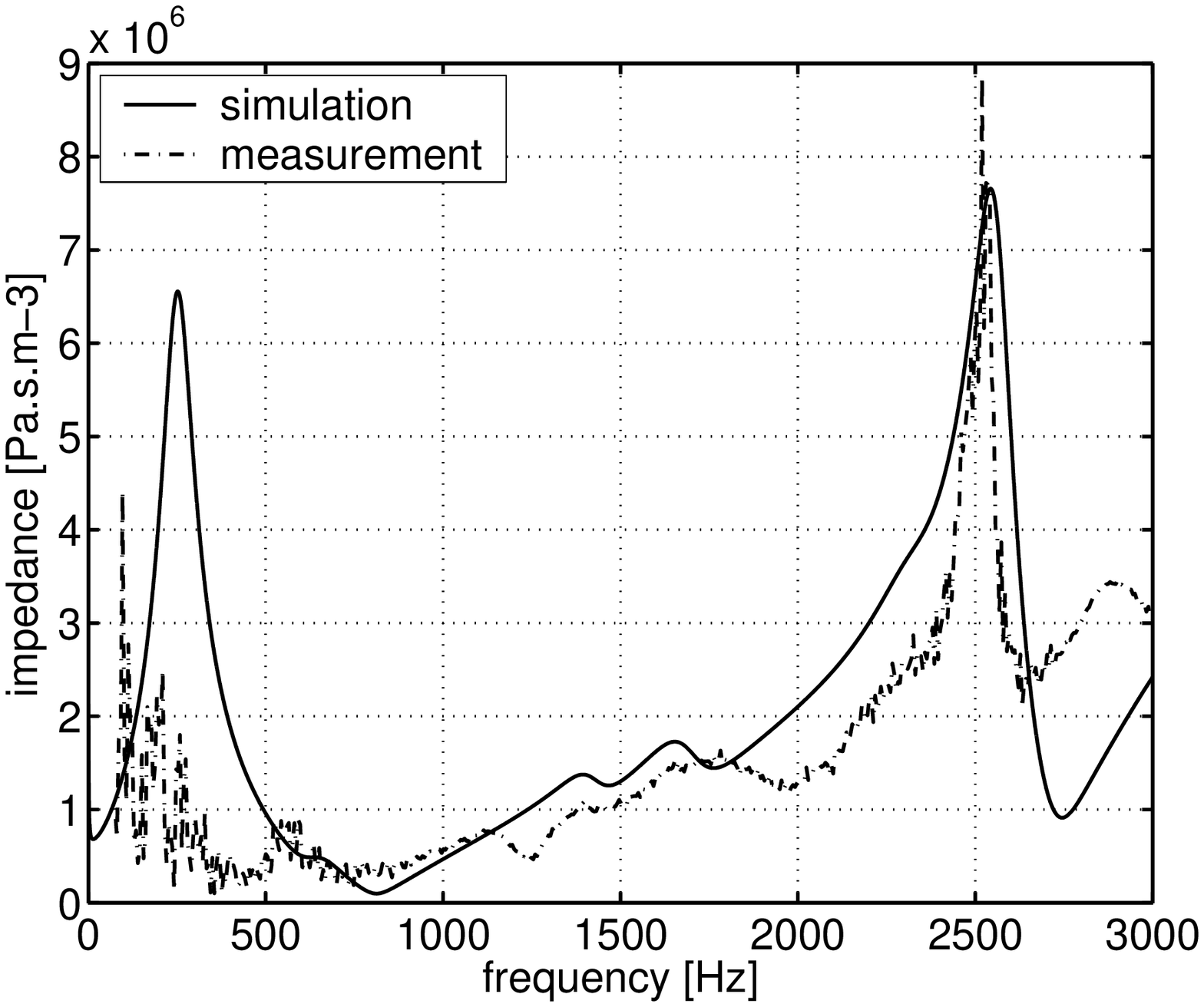}
  \caption{At left: area function for the vowel /\textopeno/ (data
      from Story and Titze with glottis added). Also 
shown are the calculated tract configuration named ``aw" by
clarinettists, as determined from inversion. The abscissa is the element
number: the first two are the glottis (see~\ref{tubes}), the next 44,
of length 4 mm  comprise a tract 174 mm long.  At right: the impedance
spectrum for the ``aw" tract configuration measured on player B.}
     \label{fig:aw}
   \end{figure}

 For the case of ``aw" (FIG.~\ref{fig:aw}) the area function found by
inversion closely resembles that of the corresponding vowel. However,
the magnitude of the first peak is too great, even allowing for
the fact that  the peak for this musician is systematically lower than
that of other musicians. Thus for the same vowel, the first peak
of player G has an amplitude between 5 and 7~MPa.s.m$^{-3}$, and is
thus  better predicted by the numerical simulation.

 Overall, the inversion results have only
moderate  agreement, and the area functions must be adjusted ``by hand''
in  order to give impedance spectra close to those measured. Further,
some  effects have been neglected, such as the fact that some players
place  the tip of the tongue just behind the lower lip, which might
plausibly  add a parallel compliance associated with the air volume
under the  tongue. Nevertheless, this does not prevent the obtaining
of  approximate area functions and allows in particular the
determination of a palatal constriction.

\subsection{Influence of the glottis opening}

\begin{figure}[h!]
  \centering
  \includegraphics[height=5.5cm, width=7cm]{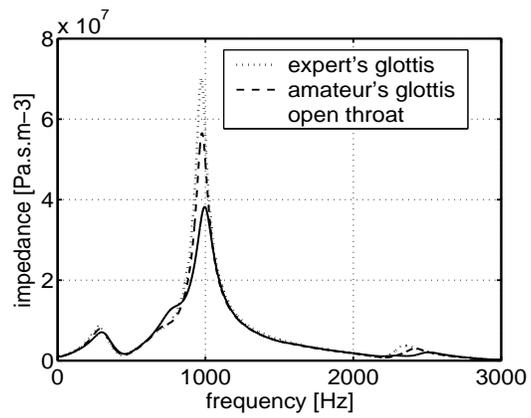}
\caption{Calculated impedance spectra for a tract in the configuration
  ``ee'' with three different values of the glottal opening. Two are
  taken from the data of Mukai for an expert and an amateur player,
  the third is for an open glottis.}\label{fig:infl-glottis}
\end{figure}

FIG.~\ref{fig:infl-glottis} shows the impedance for an ``ee''
configuration (as described in FIG.~\ref{fig:ee}) calculated in three cases:
\begin{itemize}
\item the glottis is an expert's glottis, almost closed, i.e. a slit
  of area 15 mm$^2$;
\item the glottis is an amateur's glottis, partially closed, of
  section 90 mm$^2$;
\item the glottis is the same as the previous case and the section of
  the last two cylinders of the vocal tract (just above the glottis)
  was increased for more realism: it is indeed very likely that the
  amateur not only open wider the glottis but as well the upper part
  of the vocal tract.
\end{itemize}

This figure shows that the opening of the glottis can have a large
effect (a factor of 2 or 6 dB) on the amplitude of the peaks. Some
important levels in the impedance data allow us thus to assume that in
many cases, the glottis is almost closed.

\section{CONCLUSION}   

The newly configured spectrometer permitted the measurement of the
impedance spectra of the vocal tracts of experienced clarinet players in a
situation that allowed them to mime the conditions of playing. In
contrast with most previous measurements, the players could blow into the
mouthpiece and, probably as a consequence of this, the impedance
spectra showed the strong resonances that are characteristic of a
nearly closed glottis, which is the case both for speech and for the
playing of experience wind instrument players~\cite{Muk92}. As the
glottal opening could not be monitored, this deduction is only made on both
the good reproducibility which ensures us that the glottis is well controlled
by the musician during the measurement and the high level of the
impedance in comparison to previous measurements.

   The peak values of impedance measured were in some cases comparable
   with the peak values of that of the clarinet (Wolfe \emph{et
   al.}~\cite{webUNSW}, Backus~\cite{Bac74}). Moreover, the vocal
 tract impedance is much 
   larger than the clarinet impedance around the even harmonics. The
   phase of these harmonics, when we consider the whole impedance
   (i.e. the sum of the clarinet impedance and the vocal tract one) is
   thus shifted, which may affect the playing frequency.  This
   suggests that the acoustic effects of the vocal tract should not be
   neglected and that they may have a musically significant influence
   on the sound produced.   

The combination of these measurement with a survey about the
utilisation by clarinet players of their vocal tract allow us to
relate observed acoustical responses to the reported embouchures of
the players. All players agreed that the vocal tract had a large
influence on the sound, but their opinions regarding the best
configuration to adopt differ considerably. This is a potentially
important conclusion for musicians: many highly 
respected professional clarinettists achieve fine sound quality using
rather different tract configurations. Nevertheless, two general
trends can be observed. The players try to keep their configuration
stable for most part of the register, which is in contrast to
Johnson's suggestion~\cite{Joh86} that players may tune one of the vocal
tract resonances to the note played. On the other hand, the
configuration can be changed substantially for special effects such as
difficult slurs across registers or pitch bend: players lower the
tongue and the overall magnitude of the impedance when they aim to
bend the pitch down, or to slur downwards over registers, and vice
versa.  To examine this phenomenon in more detail, we hope that, in
the future, it may be possible to make such measurements in real time
in order to determine how the musician changes his configuration during
a transition.    

\section*{ACKNOWLEDGMENTS}  

We are grateful to John Smith and David Bowman for ACUZ program and to
the Australian Research Council for funding. We would like to thank
also Brad Story for his help with the numerical model.


\begin{thebibliography}{10}

\bibitem{Bac85}
J. Backus, ``The effect of the player's vocal tract on woodwind instrument
  tone,'' J. Acoust. Soc. Am. {\bf 78}(1),  (1985).

\bibitem{Ben83}
A. Benade, ``Air column, reed and player's windway interaction in musical
  instruments,''  in {\sl Vocal Fold Physiology}, edited by I. Titze and R.
  Scherer (The Denver Center for the Performing Arts, Denver Colorado, 1983).

\bibitem{Hoe86}
P. Hoekje,  ``Intercomponent energy exchange and upstream/down\-stream symmetry
  in nonlinear self-sustained oscillations of reed instruments,'' Ph.D. thesis,
  CaseWestern Reserve University, Cleveland, Ohio, 1986.

\bibitem{Joh86}
R. Johnston, P. Clinch, and G. Troup, ``The role of the vocal tract resonance
  in clarinet playing,'' Acoustics Australia {\bf 14}(3),  (1986).

\bibitem{Wil96}
T. Wilson,  ``The measured upstream impedance for clarinet performance and its
  role in sound production,'' Ph.D. thesis, University of Washington, 1996.

\bibitem{Bac61}
J. Backus, ``Vibration of the reed and the air column in the clarinet,'' J.
  Acoust. Soc. Am. {\bf 33}(6),  (1961).

\bibitem{Muk92}
M.~S. Mukai, ``Laryngeal movement while playing wind instruments,''  in {\sl
  Proc. of the International Symposium on Musical Acoustics} (Tokyo,
  Japan, 1992), pp.\ 239--242.

\bibitem{webUNSW}
J. Wolfe, Clarinet Acoustics,
\url{http://www.phys.unsw.edu.au/music/clarinet}.

\bibitem{Bac74}
J. Backus, ``Input impedance curves for the reed woodwind instruments,'' J.
  Acoust. Soc. Am. {\bf 56}(4), 1266--1279, (1974).

\bibitem{Smi00}
J. Smith, C. Fritz, and J. Wolfe, ``A new technique for the rapid measurement
  of the acoustic impedance of wind instruments,''  in {\sl Proc. of the
  Seventh International Congress on Sound and Vibration}, edited by G. Guidati,
  H. Hunt, H. Heller, and A. Heiss (Garmisch-Partenkirchen, Germany,
  4-7 July 2000), Vol.~III, pp.\ 1833 -- 1840.

\bibitem{webVT}
C. Fritz and J. Wolfe, Impedance measurements of clarinet player's airway,
  \url{http://www.phys.unsw.edu.au/~jw/AirwayImp.html}.

\bibitem{Son74}
M. Sondhi, ``Model for wave propagation in a lossy vocal tract,'' J. Acoust.
  Soc. Am. {\bf 51}(6), 1070--1075 (1974).

\bibitem{Son87}
M. Sondhi and J. Schroeter, ``A hybrid time-frequency domain articulatory
  speech synthesizer,'' IEEE Trans. Acoust., Speech, Signal Processing {\bf
  35}, 955--967 (1987).

\bibitem{Sto00}
B. Story, A. Laukkanen, and I. Titze, ``Acoustic impedance of an artificially
  lengthened and constricted vocal tract,'' J. Voice {\bf 14}(4), 455--469
  (2000).

\bibitem{Mor00}
P. Morse and K. Ingard, {\sl Theoretical acoustics} (Princeton University
  Press, ADDRESS, 2000).

\bibitem{Fle95}
N. Fletcher and T. Rossing, {\sl The Physics of Musical Instruments}
  (Springer-Verlag, New York, 1995).

\bibitem{Sto96}
B. Story, I. Titze, and E. Hoffman, ``Vocal tract area functions from magnetic
  resonance imaging,'' J. Acoust. Soc. Am. {\bf 100}(1),  (1996).

\bibitem{ElMas98}
S. El-Masri, X. Pelorson, P. Saguet, and P. Badin, ``Development of the
  transmission line matrix method in acoustics applications to higher modes in
  the vocal tract and other complex ducts,'' Int. J. Numer. model {\bf 11},
  133--151 (1998).

\end{thebibliography}

\newpage
\listoftables
\listoffigures

\newpage

\newpage

\newpage

\newpage

\newpage

\newpage

\newpage

\newpage

\newpage

\newpage

\newpage

\newpage

\end{document}